\documentclass[showpacs,amsmath,amssymb,pra,superscriptaddress]{revtex4}

\usepackage{graphicx}
\usepackage{dcolumn}
\usepackage{bm}

\begin{document}


\title{Beryllium in Strong Magnetic Fields}
\date{\today}

\author{Omar-Alexander Al-Hujaj}
\email{Alexander.Al-Hujaj@pci.uni-heidelberg.de}
\affiliation{%
Theoretische Chemie, Institut f\"ur Physikalische Chemie der Universit\"at Heidelberg, INF 229, 69120 Heidelberg, Germany}%
\author{Peter Schmelcher}
\email{Peter.Schmelcher@pci.uni-heidelberg.de}
\affiliation{%
Theoretische Chemie, Institut f\"ur Physikalische Chemie der Universit\"at Heidelberg, INF 229, 69120 Heidelberg, Germany}%
\affiliation{%
Physikalisches Institut der Universit\"at Heidelberg, Philosophenweg 12, 69120 Heidelberg, Germany}%

\date{\today}

\begin{abstract}
We investigate the electronic structure of the beryllium atom subjected to a strong
magnetic field in the regime $0 \le \gamma \le 10$~a.u. The ground as well as many
excited states of spin singlet, triplet and quintet multiplicity covering the magnetic
quantum numbers $|M|=0,1,3,6$ for both positive and negative $z-$parity are discussed
and analyzed. Total and one-particle ionization energies are presented. Transition wavelengths
as a function of the field strengths for allowed dipole transitions are provided.

\end{abstract}

\maketitle
\section{Introduction}

The past decades have seen an enormous development with respect our knowledge
of the behavior and properties of atoms exposed to strong external fields.
This holds equally for time-dependent electric as well as static electric
or magnetic fields (for reviews of the subject concerning static fields see
Refs.~\cite{Fri89,Ruder:1994_1,Ced97,Sch98,Her03}). In case of static homogeneous magnetic
fields the hydrogen atom served as a paradigm for the effects due to the combined
Coulomb and magnetic interactions \cite{Fri89}: It is one of the most fundamental
few-degrees of freedom quantum systems whose classical counterpart shows a 
transition from regularity to chaos finally exhibiting a (almost) completely chaotic phase space. 
The investigations on hydrogen in a magnetic field had major impact on a variety
of other fields such as nonlinear dynamics, semiclassics of nonintegrable systems
as well as magnetized structures in general \cite{Fri89}. On the other hand major advances in high
resolution laser spectroscopy allowed a detailed and highly instructive comparison
of theoretical results and experimental data.

Turning to astrophysics the success of the investigations on hydrogen in a strong
magnetic field is equally impressive. During the eighties comprehensive studies of
the spectrum and eigenfunctions for bound states (see Ref.\cite{Ruder:1994_1} and Refs. therein)
were performed with particular emphasis on the regime
of field strengths occuring in the atmospheres of magnetic white dwarfs $10^2$~T~$<B<10^5$~T.
Among others, the corresponding data have lead to a conclusive
interpretation of the observed spectrum of the white dwarf GrW+70$^\circ$$8247$
which was a key to our understanding of the properties of spectra of magnetic white dwarfs in general
(see e.g. Refs. \cite{Ang85,Ang78,Gre85,Wun85,Wic88}). In the nineties detailed investigations
of the continuum properties of hydrogen in strong magnetic fields have been accomplished
\cite{Mer95,Dup95}.

More-electron systems are due to the occurence of the electron-electron repulsion
that competes with both the electron-nuclear attractions and the magnetic interactions
much more complicated. It was only in the late nineties that a sophisticated electronic
structure method became available which allows to compute the bound state properties
of the helium atom exposed to a strong field \cite{Becken:1998_all}.
Subsequently \cite{Becken:1998_all}
hundred excited states and the corresponding oscillator strengths of their transitions
have been studied with the necessary accuracy in order to perform a comparison with
astrophysical observation. Employing these data strong evidence arose that the mysterious
absorption edges of the magnetic white dwarf GD229 \cite{Gre80,Sch90,Sch96},
which were for almost 25 years unexplained, are due to helium in a strong magnetic field $B
\approx 50\ 000$~T \cite{Jor98,Jor01}.
Also very recently the newly established helium data were used to analyze a number of
magnetic and suspected-magnetic southern white dwarfs \cite{Sch02,Wic00}.

Beyond the one- and two-electron atoms our knowledge on the behavior of multi-electron
atoms subjected to strong magnetic fields is very scarce. This is in contrast to the
astrophysical necessity for further informations on the spectral properties of multi-electron atoms.
The ongoing Sloan Digital Sky Survey already doubled the number of
known magnetic white dwarfs \cite{Schxx}. It is believed that heavier atoms are present in the
atmospheres of the corresponding stars due to accretion of interstellar
matter, and particularly it is expected that these objects are quite common \cite{Rei01}.

To improve the above-mentioned situation of the lack of data for multi-electron atoms
we have very recently developed a full configuration interaction approach to more-electron ($N \ge 3$)
atoms in strong magnetic fields \cite{AlH04}. As a first investigation the lithium atom
has been studied in detail \cite{AlH04}: the ground as well as many excited states of different
symmetries have been computed for a broad range of field strengths with the astrophysically
required accuracy. Thereby our knowledge on the electronic structure of the lithium atom
in a strong magnetic field has been advanced significantly (for the state-of-the-art
before this work see Refs.\cite{Neu87,Dem94,Jon96,Ivanov:1998_1,Ivanov:2000_1,Qia00,Gua01,Mor02}).

In the present work we investigate the behavior and properties of the Beryllium atom
in the complete regime $0 \le \gamma \le 10$~a.u. (following the usual convention,
the field strengths in atomic units is denoted by $\gamma$, where $\gamma=1$
corresponds to $2.355\times10^5$~T) which covers in particular the field range
of the magnetic white dwarfs. Besides the global ground state we present an analysis
of many excited states possessing a variety of different symmetries thereby multiplying
the up-to-date information on the beryllium atom in the presence of a (strong) magnetic field.
Let us comment at this point on the existing literature.
Refs.\cite{Neu87,Mor02} contain for a few field strengths
in the high field regime a discussion of the ground state (for that field strengths).
In Ref.\cite{Iva98} the field-free ground 
state has been investigated in the presence of the field and Ref.\cite{Ivanov:2000_1}
contains a discussion of the ground state in the high field regime. Both latter investigations
deal with a Hartree-Fock approach to the electronic structure. Also within Hartree-Fock
Ref.\cite{Ivanov:2001_1} contains an investigation of the global ground state for the
weak to the high field regime. The most sophisticated correlated approach up to date
is provided in Ref.\cite{Guan:2003_1}. The regime $0 \le \gamma \le 1.0$ is covered
and the ground state as well as several excited states are studied with a relatively
high accuracy thereby employing a frozen-core correlated approach. When comparing our data
with the existing ones in the literature the latter will predominantly be from Ref.
\cite{Guan:2003_1}.

We proceed as follows. Section \ref{sec:hamil} contains a discussion of the fixed-nucleus Hamiltonian
and its symmetries. Section \ref{sec:method} describes our computational approach. Section
\ref{sec:res} presents a discussion and analysis of our results. It starts with a discussion
of the global ground state of the beryllium atom with increasing field strength and turns then
to a detailed study of the excited states of various symmetries. In section \ref{sec:wave} 
we report on the transition wavelengths as a function of the field strength. Finally
section \ref{sec:summary} contains the conclusions.

\section{Hamiltonian and symmetries}
\label{sec:hamil}

The electronic Hamiltonian describing the beryllium atom exposed to a strong
magnetic field assuming an infinitely heavy nucleus reads as follows
\begin{eqnarray}
  \label{eq:ham1}
  H(\bm{B})&=&\sum_{i=1}^4 H_i(\bm{B}) + \sum_{i\not=j}H_{ij}\qquad \mbox{with}\\
  \label{eq:ham2}
  H_i(\bm{B})&=&\frac{\bm{p}_i^2}{2}+\frac{\bm{B}\cdot\bm{l_i}}{2}+\frac{(\bm{B}\times\bm{r}_i)^2}{8}-\frac{4}{|\bm{r}_i|}+\frac{g
  \bm{B}\cdot\bm{s}_i}{2}\\
  \label{eq:ham3}
H_{ij}&=&\frac{1}{2}\frac{1}{|\bm{r}_i-\bm{r}_j|}.
\end{eqnarray}
where we have adopted the symmetric gauge for the vector potential and we 
have employed atomic units. $H_i(\bm{B})$ represents the field-dependent one-particle operator
containing the Zeeman-term $1/2\ \bm{B}\cdot\bm{l_i}$,
the diamagnetic term $1/8(\bm{B}\times\bm{r}_i)^2$, the
Coulomb interaction with the nucleus $-4/|\bm{r}_i|$ as well as
the spin Zeeman contribution $g/2\  \bm{B}\cdot\bm{s}_i$. 
$H_{ij}$ is a two-particle operator and represents the electron-electron Coulomb repulsion.

In the above Hamiltonian the direction of the magnetic field coincides with the $z-$axis.
Therefore the $z-$projection of the total orbital angular momentum and of the total spin $S_z$
are conserved. These yield the good quantum numbers $M$ and $S_z$, respectively.  
Furthermore the total spin is conserved and the $z$~parity $\Pi_z$ represents a symmetry. In the
following we employ the spectroscopic notation $\nu^{2S+1}M^{\Pi_z}$ for the electronic
states, where $\nu$ stands for the degree of excitation of a certain state with given
quantum numbers $S,M,\Pi_z$. 

\section{Computational Method}
\label{sec:method}

The energies of the electronic bound states are determined by mapping the
Schr\"{o}dinger equation to an ordinary eigenvalue problem. This is
done by applying a direct full configuration interaction (full CI)
approach. For the basis set of atomic orbitals we choose anisotropic Gaussian
functions, which have been introduced for atomic and molecular
calculations in strong magnetic fields by Schmelcher and Cederbaum
\cite{Schmelcher:1988_1}. This basis set has been successfully applied to several
atoms, ions and molecules \cite{Kappes:1996_1,Detmer:1997_all,Becken:1998_all,Al-Hujaj:2000_1}.
A major advantage of it is the flexibility:
The anisotropic Gaussian functions can be adapted to any (zero or non zero)
field strength by adjusting the corresponding nonlinear variational parameters.
As a result the convergence properties of the electronic structure calculations
using the optimized basis sets (see below) are good for any field strengths, although
it turns out that for the intrinsically most complex intermediate field regime
our approach converges fastest. The above is in contrast to symmetry adapted basis sets,
that have been employed in electronic structure investigations in fields:
typically electronic eigenfunctions can be well-described for either low field strengths
(approximate spherical symmetry) or for very strong fields 
(approximate cylindrical symmetry) but lack to converge in the intermediate field regime.

To take full advantage of the flexibility of the basis functions they have to be optimized
for each field strength and each symmetry subspace separately.
To this end we determine the primitive Gaussian functions by solving different
one-particle problems (Be$^{3+}$, He$^+$ and H) for each field
strength. In order to converge rapidly thereby using a minimal number of basis functions,
the latter have to be selected very carefully.

The four-electron beryllium calculations start with the computation of the overlap
matrix $\bm{S}(m_j,\pi_{z_j})$ of the Gaussian functions. Only those eigenvectors
$\{\bm{v}_{s_j}(m_j,\pi_{z_j})\}$ of $\bm{S}$ possessing an eigenvalue above an
appropriately chosen threshold $\varepsilon$ are used in the following calculations.
As a next step the Schr\"{o}dinger equation belonging to the
one-particle Hamiltonian (\ref{eq:ham2}) is represented as an ordinary matrix eigenvalue
problem. The numerical solution of this eigenvalue problem yields the
eigenvectors $\{\bm{h}_i(m_j,\pi_{z_j})\}$, which then
serve as the spatial part of the one-particle functions for the four-electron
investigations. Spinors $\chi_j$ are constructed as a product of the usual spin
eigenfunctions  $\bm{\alpha}$ and $\bm{\beta}$ and the orthogonal
functions $\{\bm{h}_i(m_j,\pi_{z_j})\}$. Full CI calculations are
performed with four-electron Slater determinants, which are
constructed from spinors, thereby obeying the correct symmetries, i.e.
\begin{eqnarray}
  \sum_{i=1}^4 m_i&=&M \\
  \prod_{i=1}^4 \pi_{z_i}&=&\Pi_z\\
  \sum_{i=1}^4 s_{z_i}&=&S_z.
\end{eqnarray}

In general the core electrons and specifically the $1s^2$ and $2p_0$
orbitals, are well described by basis functions that are optimized for the
$1^20^+$- and $1^20^-$-states of Be$^{3+}$, respectively.  
Further basis functions for the full CI calculations of the beryllium
atom are obtained by optimizing them for the corresponding one-particle problems of
nuclear charge $Z=1$ and $Z=2$. For the spin singlet and triplet states of beryllium
exclusively only the hydrogen atom has been employed for the optimization. 
For the quintet states, some of the
low-lying orbitals were described by functions optimized for He$^+$. To
be more specific, for the $\nu^10^+$ states functions with
one-particle symmetry $m^{\pi_z}=0^+,$ $0^-,$ and $\pm1^+$ have been
employed. The same holds for the symmetry subspaces $^1(-1)^+$ and
$^3(-1)^+$. In order to achieve convergence also for the excited states,
orbitals optimized for the excited states of
the $m^{\pi_z}=0^+$ symmetry in case of $M=0$ and of the
$m^{\pi_z}=(\pm1)^+$ symmetry in case of $M=-1$ optimized for hydrogen states are important.
For the symmetry subspace $^1(-1)^-$ we used more functions of the
one-particle symmetry  $0^-$ optimized for the ground and excited states
of hydrogen. Additionally states for $m^{\pi_z}=(\pm1)^\pm$ have been used. For the
triplet and quintet symmetry subspace with positive $z$-parity and
$M=-3$ the orbitals of $m^{\pi_z}=(\pm2)^+$ replace those with
$(\pm1)^-$ and $0^-$. For the quintet symmetry subspaces $^5(-3)^-$
and $^5(-6)^+$ additionally functions of $m^{\pi_z}=(\pm3)^+$ symmetry character
that have been optimized for hydrogen are included in the calculations. 
Our calculations have been performed using up to 50 one-particle basis
functions resulting in $80\ 000$ Slater determinants.

\section{Energies}
\label{sec:res}

\subsection{The global ground state and the properties of total energies}
\label{sec:glob_ground}

\begin{figure}[htbp]
  \centering
  \includegraphics[scale=0.3]{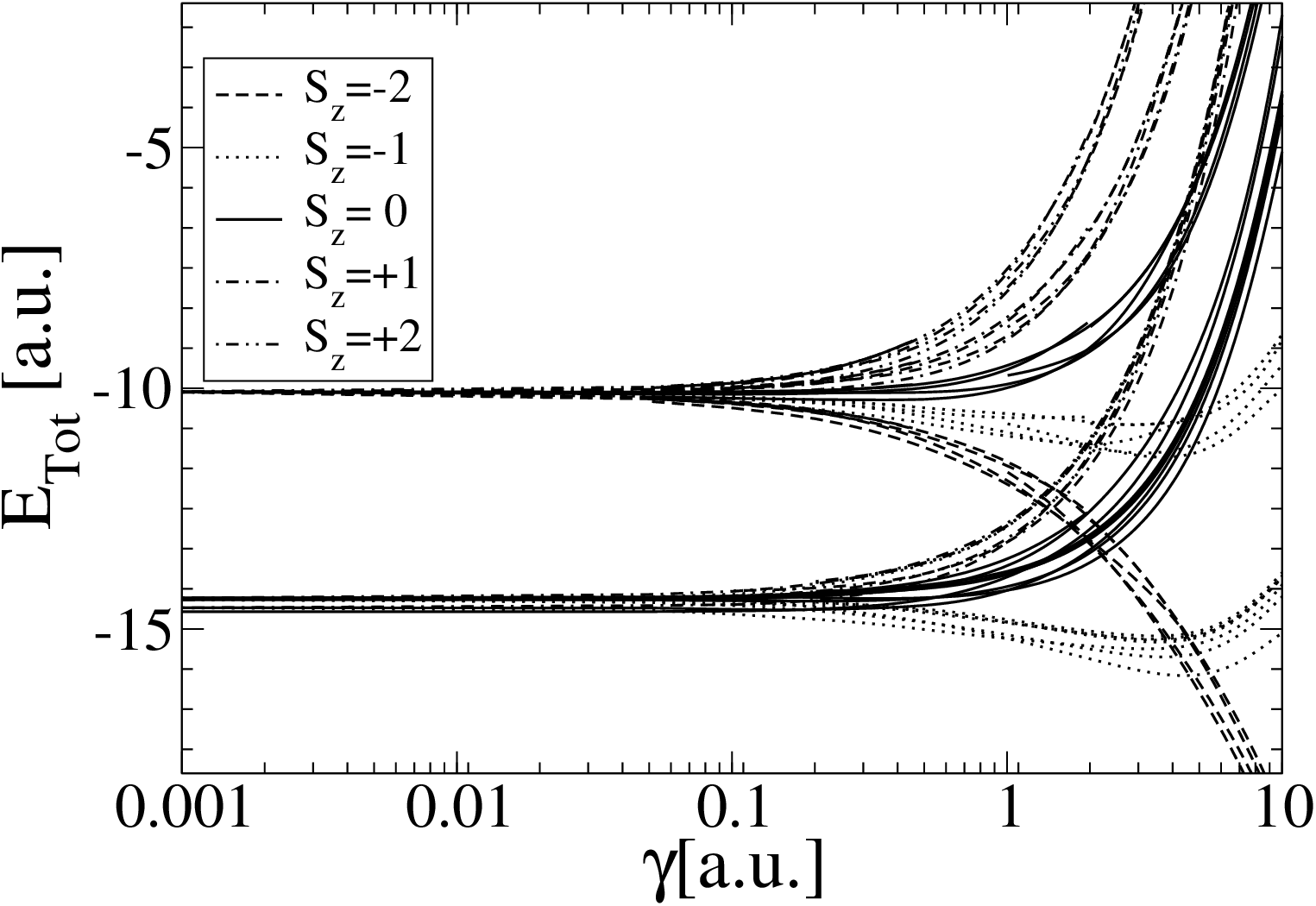}
  \caption{Total energies E$_{\mbox{\scriptsize Tot}}$ of all 61 calculated
  bound states with non positive magnetic quantum number $M$ in a.u. as a function of the
  magnetic field strength $\gamma$. It can be seen that the behavior of the
  total energies can be classified by the spin projection $S_z$.} 
  \label{fig:alltotal_be}
\end{figure}

In strong magnetic fields, more precisely in the high field regime,
the total energies of bound states of an
atomic or molecular system are dominated by the orbital and
spin Zeeman-contributions on the one hand side and the increase of the kinetic
energy on the other hand side. The kinetic energy 
depends approximately linear on the field strength (Landau zero point energy of the electrons)
and the contributions of the Zeeman terms behave exactly linear with the
changing magnetic field strength. The overall effect of these contributions on the
total energies are demonstrated in Fig.~\ref{fig:alltotal_be}.
The latter shows the total energies of all 61 calculated bound states possessing negative
magnetic quantum numbers as a function of the field strength $\gamma$. In the high 
field limit a general pattern can be identified: Energy
levels corresponding to fully spin polarized 
states ($S_z=-2$) decrease monotonically with increasing field strength,
energy levels belonging to states with $S_z=-1$ pass
through a local minimum, and those belonging to states with a positive total spin
projection increase monotonically. This classification of the overall behavior
of the energy curves depends exclusively on the spin projection of the states considered.

\begin{figure}[htbp]
  \centering
  \includegraphics*[scale=0.3]{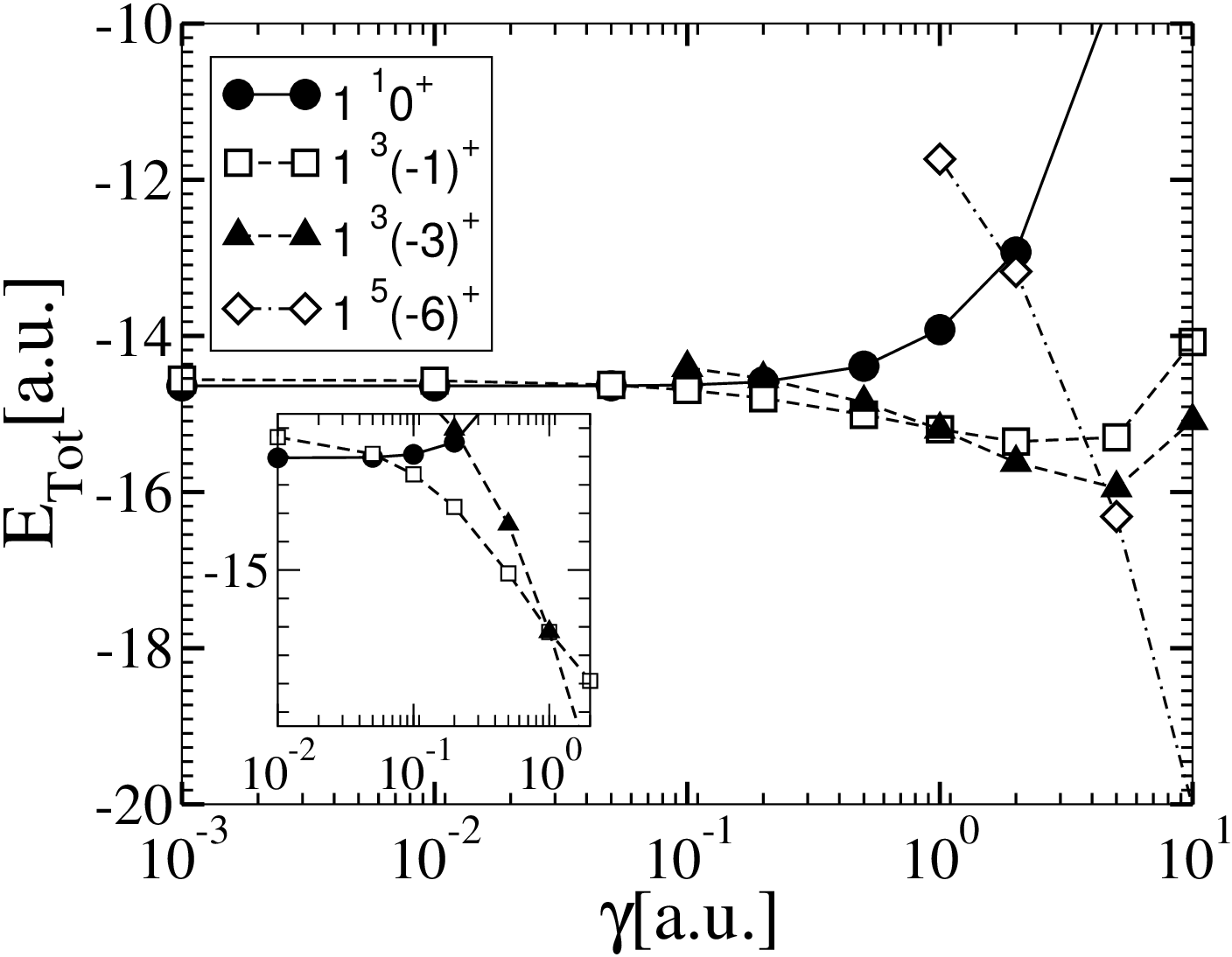}
  \caption{Total energies $E_{\mbox{\scriptsize Tot}}$ of the states constituting
  the ground state of the beryllium atom in a.u. as a function of the magnetic field
  strength $\gamma$. The inset shows a magnification for a certain regime of
  field strengths.}
  \label{fig:groundstate_be}
\end{figure}

With increasing field strength the ground state of an atom exposed to a magnetic
field changes its symmetry i.e. a crossover of several states individually representing the ground
state of the atom for a certain interval of field strengths takes place.
The orbital and spin Zeeman-terms are important for the identification
of the field-dependent global ground state 
\cite{Ivanov:1998_1,Ivanov:1999_1,Ivanov:2000_1,Ivanov:2001_1,Ivanov:2001_2}.
Beyond this the magnetically tightly bound orbitals
\cite{Loudon:1959_1,Ruder:1994_1} compared to the non-tightly bound
orbitals/states play a central role for the determination of the ground state.
Since the ground state is finally obtained by a complicated 
interplay of the above contributions, the electronic states forming
the ground state of the atom depending on the field strength
are not predictable \emph{a priori}, but have to be
determined by explicit electronic structure calculations. 
For the beryllium atom investigated here, four electronic states representing the global
ground state for different field regimes have been found. This result coincides with an
earlier one \cite{Ivanov:2001_1} based on Hartree-Fock calculations of the beryllium atom.
Having fully correlated results at hand it is now, however, possible to finally 
conclude on the ground state properties of the atom. The total energies of the states 
constituting the ground state are presented in Fig.~\ref{fig:groundstate_be}.
The low field ground state $0<\gamma<0.0612$ is the highly correlated, doubly tightly bound
singlet state $1^10^+$. Its total energy increases monotonically as can be observed
for all states with a vanishing spin quantum number $S_z=0$. 
In the field regime $0.0612<\gamma<1.0$ the global ground 
state is represented by the triply tightly bound triplet state
$1^3(-1)^+$. Concerning the contributions to its total energy the increase of the
kinetic energy is now partly compensated by the spin Zeeman-term. Therefore the energy of
this state does not monotonically increase with increasing field strength, but passes through a
minimum, which is located at $\gamma\approx3.22$. For
$1.0<\gamma<4.62$ the ground state is given by the triplet state
$1^3(-3)^+$ which possesses four tightly bound orbitals. Similar to the
$1^3(-1)^+$ state, the reader should note, that its total energy passes
through a minimum, which, for this state, is located at a higher field
strength, compared to the state $1^3(-1)^+$, namely at $\gamma\approx 4.23$. In the high
field regime $\gamma>4.23$ the spin Zeeman-term completely dominates
the total energies. The high field ground state is
provided by the quadruply tightly bound state $1^5(-6)^+$ of spin quintet symmetry. Its
energy monotonically decreases as a function of the field strength
$\gamma$. In the field-free case this state is unbound.

Since the Zeeman energy and the kinetic energy predominate the total energy of the states
and mask the interesting new properties of the atomic states, we will in the following
concentrate our discussion on the ionization energies. These will reveal
the binding properties of the corresponding states. 
For each symmetry subspace and each field strength, the corresponding ionization
threshold has to be identified. We define here the ionization energy to respect
the symmetry properties: The threshold $E_T(\gamma,M,S_z)$ for
total magnetic quantum number $M$ and $z$~projection $S_z$
of the total spin is defined by the  equation
\begin{equation}
  E_T(\gamma,M,S_z)=\min_{M_1,S_{z_1}} E^{\mbox{\scriptsize
  Be}^+}(\gamma,M_1,S_{z_1})+E^{\mbox{\scriptsize e}^-}(\gamma,M_2,S_{z_2}),
\end{equation}
where $E^{\mbox{\scriptsize Be}^+}(\gamma,M_1,S_{z_1})$ and
$E^{\mbox{\scriptsize e}^-}(\gamma,M_2,S_{z_2})$ are the total energies of
the Be$^+$ ion and the electron, respectively, depending on their
magnetic quantum numbers $M_i$ and $S_{z_i}$ ($i=1,2$). 
Obviously this requires that the electronic quantum numbers $M_2$ and
$S_{z_2}$ are related to the ionic respectively atomic quantum numbers via
\begin{equation}
  M_2=M-M_1 \qquad S_{z_2}=S_z-S_{z_1}.
\end{equation}
In order to determine the ionization thresholds many  states of the Be$^+$
ion have to be computed as a function of the magnetic field strengths $\gamma$ in
detail. Table~\ref{tab:be+} presents our results 
for the total energies of the electronic states of Be$^+$, which are associated
to the one-particle ionization thresholds. 

\begingroup
\begin{table}[ht]
\begin{ruledtabular}\begin{tabular}{c*{5}{c}}
\rule[-0.7em]{0mm}{2.0em} &\multicolumn{1}{c}{$1 ^20^+$} & \multicolumn{1}{c}{$1 ^20^-$} &\multicolumn{1}{c}{$1 ^2(-1)^+$} & \multicolumn{1}{c}{$1 ^4(-1)^+$}
& \multicolumn{1}{c}{$1 ^4(-3)^+$}\\\hline\hline
\multicolumn{1}{c}{\rule[-0.7em]{0mm}{2.0em}$\gamma $} & \multicolumn{1}{c}{$E_{\mbox{\scriptsize tot}}$ } & \multicolumn{1}{c}{$E_{\mbox{\scriptsize tot}}$ }& \multicolumn{1}{c}{$E_{\mbox{\scriptsize tot}}$ } & \multicolumn{1}{c}{$E_{\mbox{\scriptsize tot}}$ } & \multicolumn{1}{c}{$E_{\mbox{\scriptsize tot}}$ }\\
\hline
 0.000  & -14.3247  & -14.1727 &  -14.1741 & -10.0650 & -9.4156   \\
 0.001  & -14.3251  & -14.1720 &  -14.1751 & -10.0655 & -9.4185   \\
 0.010  & -14.3296  & -14.1765 &  -14.1841 & -10.0827 & -9.4452   \\
 0.050  & -14.3482  & -14.1955 &  -14.2216 & -10.1607 &  \\
 0.100  & -14.3694  & -14.2180 &  -14.2672 & -10.2575 & -9.6888   \\
 0.200  & -14.4038  & -14.2572 &  -14.3476 & -10.4353 & -9.9243   \\
 0.500  & -14.4606  & -14.3448 &  -14.5358 & -10.8918 & -10.5188  \\
 1.000  & -14.4630  & -14.4255 &  -14.7520 & -11.4967 & -11.3203  \\
 2.000  & -14.3300  & -14.4558 &  -15.0000 & -12.4090 & -12.6002  \\
 5.000  & -13.5971  & -13.9795 &  -15.0184 & -14.5106 & -15.4367  \\
10.000  & -11.6231  & -12.1626 &  -13.8087 & -17.2291 & -18.8283  \\
\hline
\end{tabular}
\end{ruledtabular}
\caption{Total energies for Be$^+$ in a.u. needed to determine the one-particle
  ionization thresholds for different field strengths for the 
  beryllium states considered here.\label{tab:be+}}
\end{table}
\endgroup

One implication of the above definition of the ionization threshold is the
fact, that the ionization energy neither depends on the value of the
spin projection $S_z$, nor on the sign of the magnetic quantum
number. Therefore, total energies will be provided for one spin
projection and the non-positive magnetic quantum number. The maximal negative value of the spin
projection onto the magnetic field axis is chosen, i.e $S_z=-S$. By adding the
corresponding contributions of the spin Zeeman- and orbital Zeeman-term
respectively, the total energies for the states with different quantum
numbers $S_z$ and also for the opposite sign of the orbital quantum number $M$ can be obtained.

\subsection{The symmetry subspace $^10^+$.}
\label{sec:10+}

\begin{figure}[htbp]
  \includegraphics*[scale=0.30]{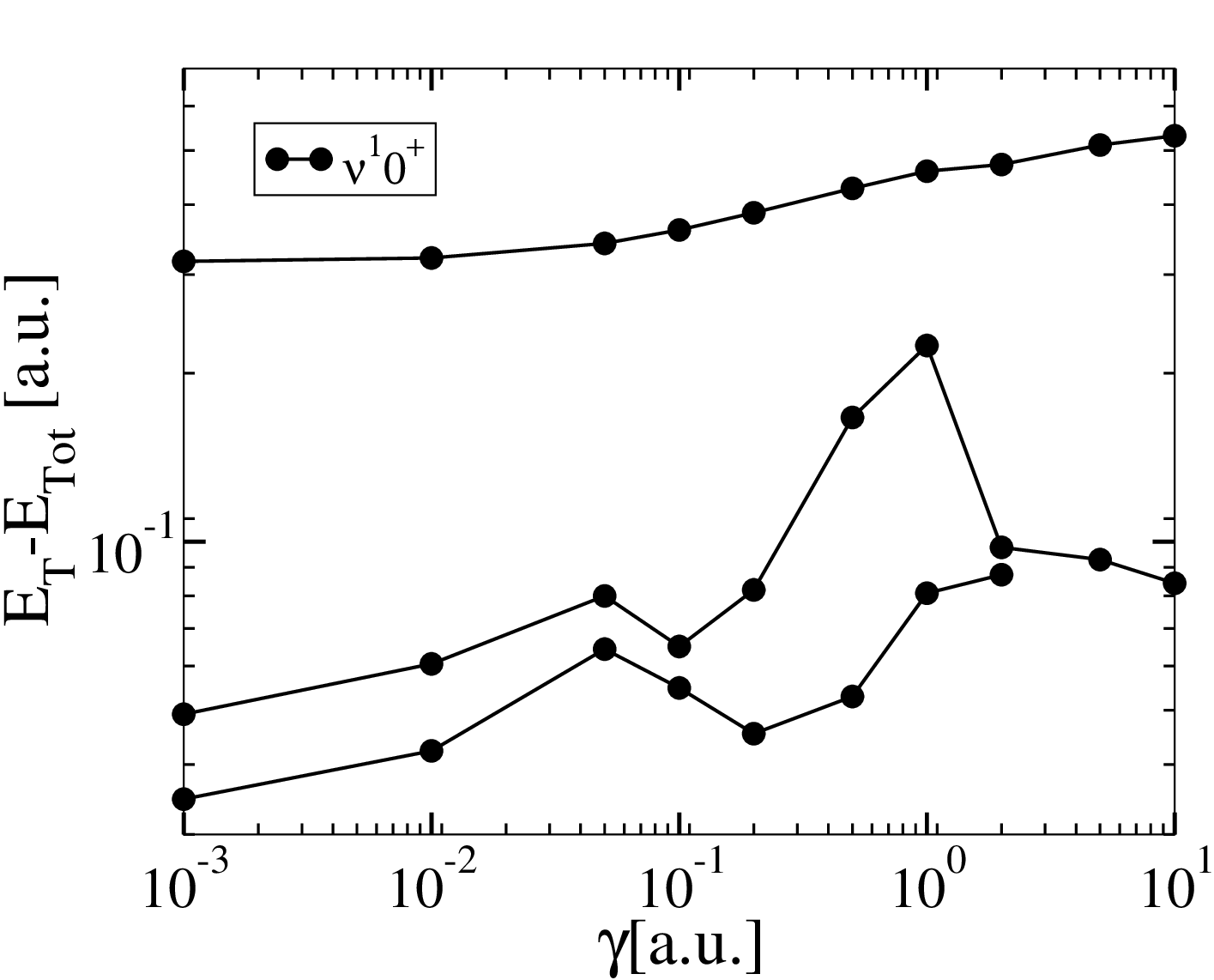}
  \caption{One-particle ionization energies for the states $\nu^10^+$ ($\nu=1,2,3$) in a.u. as a function of the magnetic field strength $\gamma$.}
  \label{fig:10+}
\end{figure}

The ionization energies for the $\nu^10^+$ states
($\nu=1,2,3$) are presented in Fig.~\ref{fig:10+}. Numerical values for the
total and ionization energies, the symmetries of the ionization threshold,
as well as a comparison to previously published data can be found in
table~\ref{tab:10+}. The state $1^10^+$ represents, as mentioned
above, the ground state of the atom in the low field regime. Its
ionization energy increases monotonically with increasing field
strength. At $\gamma=0$ it is $0.3158$~a.u., whereas for $\gamma=10$ it
has increased to $0.5311$~a.u. Comparing the total energies obtained for the $1^10^+$ state
within our fully correlated CI approach with the existing data in the literature we arrive at the
following conclusions. In the absence of the external field the relative accuracy
is $2 \times 10^{-3}$ compared to the more accurate calculations in Ref.(\cite{Galvez:2002_1}).
This holds equally for the comparison of our data for finite field strengths
with the more accurate results in Ref.(\cite{Guan:2003_1})
which uses a frozen-core correlated approach.
The latter statement is correct up to $\gamma \approx 0.5$ whereas for $\gamma > 0.5$ our method
yields lower total energies than the one employed in Ref.(\cite{Guan:2003_1}).
For field strengths $\gamma \ge 2$ only Hartree-Fock results are available for
comparison \cite{Ivanov:1998_1,Ivanov:2001_1} and it turns out that these are energetically
higher up to one percent for $\gamma = 10$.

For the first excited state $2^10^+$ a different behavior can be observed.
Its one-particle ionization energy (OPIE)
increases for low fields $0<\gamma<0.05$. In the field range
$0.05<\gamma<0.2$ it passes through a local minimum and increases
from $0.0650$~a.u. for $\gamma=0.1$ to $0.2241$ at
$\gamma=1$. For $\gamma \le 1$ the ionization threshold involves the $1^20^+$ electronic state
of Be$^+$ whereas for $\gamma > 1$ it is the $1^20^-$ state. As a consequence, the ionization
threshold for the $2^10^+$ state drops to  $0.0977$~a.u. at
$\gamma=2$. For $\gamma>2$ the ionization energy decreases as a
function of $\gamma$ at a much lower rate. Only for $\gamma = 0$ 
energies of this state are available for comparison \cite{Galvez:2002_1} i.e. this state has not
been investigated previously in the literature. It turns out that our energies show a
relative accuracy of $2 \times 10^{-3}$ similar to the case of the $1^10^+$ state in the 
absence of the field. 

\begingroup
\begin{table*}[ht]\squeezetable
\begin{ruledtabular}
\begin{tabular}{*{3}{c}*{9}{c}}
\multicolumn{2}{c}{\rule[-0.7em]{0mm}{2.0em}} &\multicolumn{3}{c}{\rule[-0.7em]{0mm}{2.0em}{}$1^10^+$}& \multicolumn{3}{c}{$2 ^10^+$}& \multicolumn{3}{c}{$3 ^10^+$}\\
\hline\hline
\multicolumn{1}{c}{\rule[-0.7em]{0mm}{2.0em}$\gamma $}&\multicolumn{1}{c}{T$_{\mbox{\scriptsize{}Sym}}$}  & \multicolumn{1}{c}{E$_{\mbox{\scriptsize{}tot}}$} &\multicolumn{1}{c}{E$_{\mbox{\scriptsize{}Ion}}$}  & \multicolumn{1}{c}{E$_{\mbox{\scriptsize{}Lit}}$}   & \multicolumn{1}{c}{E$_{\mbox{\scriptsize{}tot}}$}&  \multicolumn{1}{c}{E$_{\mbox{\scriptsize{}Ion}}$} &\multicolumn{1}{c}{E$_{\mbox{\scriptsize{}Lit}}$}& \multicolumn{1}{c}{E$_{\mbox{\scriptsize{}tot}}$} &\multicolumn{1}{c}{E$_{\mbox{\scriptsize{}Ion}}$}&  \multicolumn{1}{c}{E$_{\mbox{\scriptsize{}Lit}}$} \\
\hline
 0.000  & $     ^20^+$  &-14.6405  &0.3158    &-14.66736\footnotemark[1]& -14.3908  &0.0661    & -14.41824\footnotemark[1] & -14.3578  &0.0331     &-14.40793\footnotemark[1] \\
 0.001  & $     ^20^+$  &-14.6410  &0.3168    &-14.66287\footnotemark[2]& -14.3733  &0.0492    &                           & -14.3589  &0.0347     &\\              
 0.010  & $     ^20^+$  &-14.6408  &0.3212    &-14.66279\footnotemark[2]& -14.3801  &0.0605    &                           & -14.3619  &0.0423     &\\              
 0.050  & $     ^20^+$  &-14.6393  &0.3411    &-14.56986\footnotemark[3]& -14.3782  &0.0800    &                           & -14.3625  &0.0643     &\\              
 0.100  & $     ^20^+$  &-14.6298  &0.3604    &-14.64955\footnotemark[2]& -14.3344  &0.0650    &                           & -14.3241  &0.0548     &\\              
 0.200  & $     ^20^+$  &-14.5907  &0.3868    &-14.61160\footnotemark[2]& -14.2858  &0.0820    &                           & -14.2492  &0.0454     &\\              
 0.500  & $     ^20^+$  &-14.3882  &0.4276    &-14.40818\footnotemark[2]& -14.1273  &0.1667    &                           & -14.0136  &0.0529     &\\              
 1.000  & $     ^20^+$  &-13.9220  &0.4590    &-13.91717\footnotemark[2]& -13.6871  &0.2241    &                           & -13.5439  &0.0809     &\\              
 2.000  & $     ^20^-$  &-12.9275  &0.4717    &-12.88908\footnotemark[3]& -12.5535  &0.0977    &                           & -12.5430  &0.0873     &\\              
 5.000  & $     ^20^-$  &-9.4907  &0.5112     &-9.40602\footnotemark[3] & -9.0724   &0.0929    &   \\  
10.000  & $     ^20^-$  &-2.6936  &0.5310     &-2.5988\footnotemark[4]  & -2.2469   &0.0843    &     \\
\end{tabular}
\caption{Total energies E$_{\mbox{\scriptsize Tot}}$,  one-particle
  ionization energies E$_{\mbox{\scriptsize Ion}}$, and
previously published data for the total energies E$_{\mbox{\scriptsize Lit}}$ in a. u. for the states $\nu^10^+$ ($\nu=1,2,3$), as well as the symmetry T$_{\mbox{\scriptsize{}Sym}}$ of the $Be^{+}$ ion belonging to the ionization threshold for different field strengths
$\gamma$.}\label{tab:10+} \footnotetext[1]{Ref. \cite{Galvez:2002_1}.} \footnotetext[2]{Ref. \cite{Guan:2003_1}.}
  \footnotetext[3]{Ref. \cite{Ivanov:2001_1}.} \footnotetext[4]{Ref. \cite{Ivanov:1998_1}.}
\end{ruledtabular}
\end{table*}
\endgroup

The OPIE of the second excited state $3^10^+$ behaves as a function of
the field strength similar to the first excited state.
It increases for low fields $0<\gamma<0.05$ and passes through a local minimum at
$\gamma\approx0.2$ For $0.2>\gamma>2$ it increases. For $\gamma>2$ we
were not able to obtain sufficiently accurate results.

\subsection{The symmetry subspaces $^{1/3}(-1)^+$.}
\label{sec:1-1+}

\begin{figure}[htbp]
  \includegraphics*[scale=0.30]{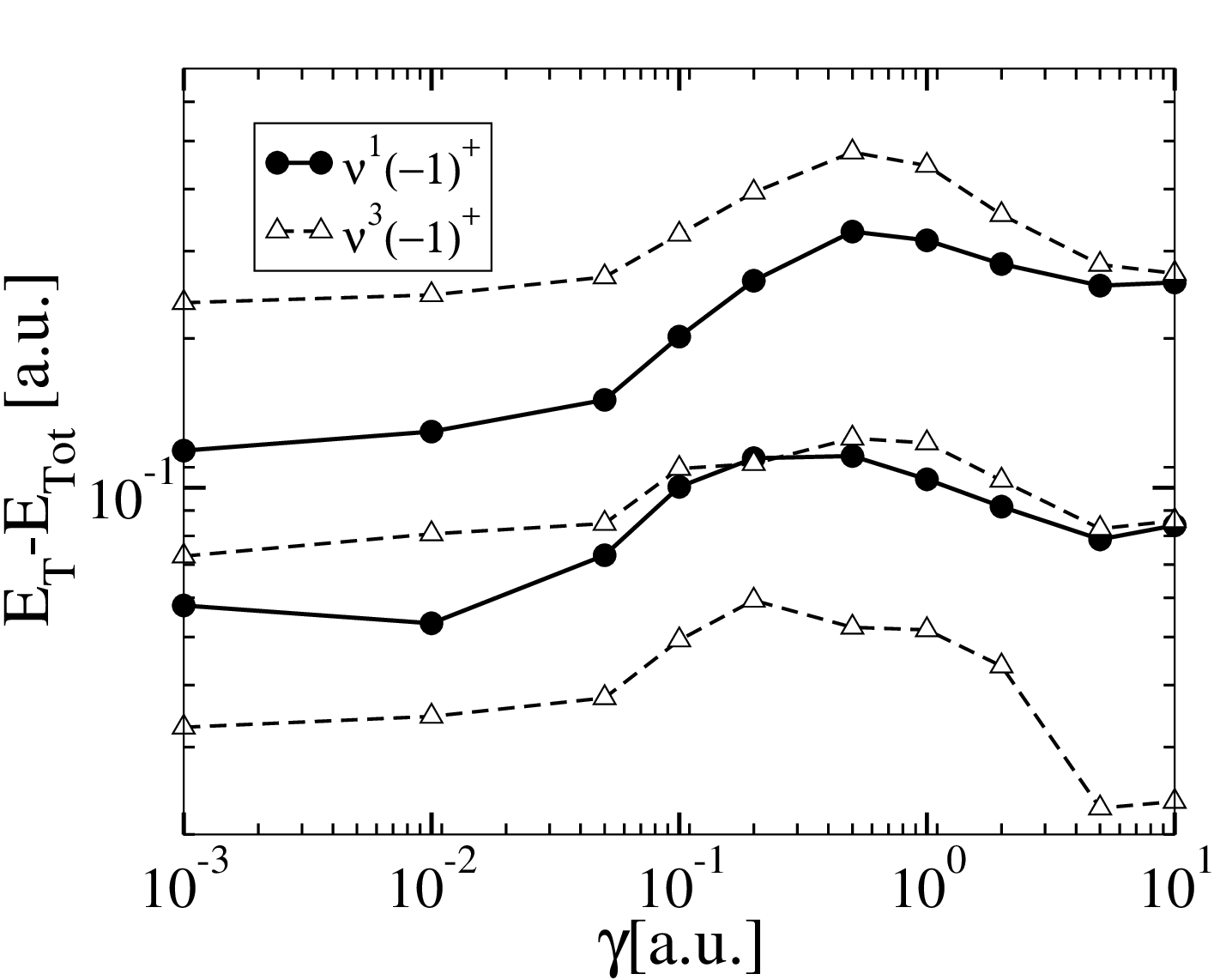}
  \caption{One-particle ionization energies for the states
    $\nu^{1}(-1)^+$ ($\nu=1,2$) and $\nu^{3}(-1)^+$ ($\nu=1,2,3$)  in
    a.u. as a function of the magnetic field strength $\gamma$.} 
  \label{fig:2S+1(-1)+}
\end{figure}

This subsection presents a discussion of our results on the singlet states $\nu^1(-1)^+$ 
for $\nu=1,2$ and the triplet states $\nu^3(-1)^+$ for $\nu=1-3$. We
illustrate the corresponding ionization energies in Fig.~\ref{fig:2S+1(-1)+} and the
numerical values for the total energies, OPIEs
together with previously published data for the
total energies of these states are provided in table~\ref{tab:1-1+}
and in table~\ref{tab:3-1+}, respectively. The symmetries of the corresponding
ground electronic states of the Be$^{+}$ ion in the ionization limit are also provided
as a function of the field strength.

Let us first discuss the behavior of the energies of the singlet states.
The OPIEs of the energetically lowest state $1^1(-1)^+$
which is a triply tightly bound state, does not increase monotonically
but passes through a local maximum at $\gamma\approx 0.5$.
The increase of the OPIE in the low field regime
$0<\gamma<0.5$ is more pronounced for the $1^1(-1)^+$ state, than for
the $1^10^+$. It increases approximately by a factor 3 from $0.1161$~a.u. at $\gamma=0$ to
$0.3284$~a.u. at $\gamma=0.5$. For $\gamma>0.5$ the
OPIE decreases. One reason for this is the field-dependent
behavior of the ionization threshold. For $\gamma>0.2$ it involves
the triply tightly bound Be$^+$ state $1^1(-1)^+$. This threshold has
in the latter field regime a lower energy than the threshold involving
the ionic state $1^20^-$, which represents the
high field ionization threshold for the $^10^+$ symmetry subspace.
Considering the numerical values for the total energies of the $1^1(-1)^+$ state
compared to the literature we arrive at the following conclusions. For the 
field-free case the relative accuracy is approximately $2 \times 10^{-3}$ in
comparing with the highly accurate values presented in Ref.\cite{Galvez:2002_1}.
In the presence of the field for $\gamma \le 0.5$ the comparison with
Ref.\cite{Guan:2003_1} shows that our values are again by a relative factor of
$2 \times 10^{-3}$ above the values of this reference. However, for $\gamma > 1$
no data are available up to date in the literature.

The OPIE of the first excited state
$2^1(-1)^+$ shows a similar behavior as that of the ground state.
It increases in the low field regime and decreases
in the high field regime thereby exhibiting a local maximum at 
$\gamma \approx 0.2$. Total energies for this state are available
only for a vanishing external field and our computed value shows in
this case a relative accuracy of $10^{-3}$ compared to the value
provided in Ref.\cite{Galvez:2002_1}.

\begin{table*}[ht]
\begin{ruledtabular}
\begin{tabular}{*{3}{c}*{6}{c}}
\multicolumn{2}{c}{\rule[-0.7em]{0mm}{2.0em}} &\multicolumn{3}{c}{\rule[-0.7em]{0mm}{2.0em}{}$1^1(-1)^+$}& \multicolumn{3}{c}{$2 ^1(-1)^+$}\\
\hline\hline
\multicolumn{1}{c}{\rule[-0.7em]{0mm}{2.0em}$\gamma $}&\multicolumn{1}{c}{T$_{\mbox{\scriptsize{}Sym}}$}  & \multicolumn{1}{c}{E$_{\mbox{\scriptsize{}tot}}$} &\multicolumn{1}{c}{E$_{\mbox{\scriptsize{}Ion}}$}  & \multicolumn{1}{c}{E$_{\mbox{\scriptsize{}Lit}}$}   & \multicolumn{1}{c}{E$_{\mbox{\scriptsize{}tot}}$}&  \multicolumn{1}{c}{E$_{\mbox{\scriptsize{}Ion}}$} &\multicolumn{1}{c}{E$_{\mbox{\scriptsize{}Lit}}$} \\
\hline
 0.000  & $     ^20^+$  &-14.4408  &0.1161 &-14.47344\footnotemark[1]  & -14.3798  &0.0551   & -14.39311\footnotemark[1]  \\
 0.001  & $     ^20^+$  &-14.4429  &0.1188 &-14.46849\footnotemark[2]  & -14.3820  &0.0579                                \\
 0.010  & $     ^20^+$  &-14.4494  &0.1298 &-14.47276\footnotemark[2]  & -14.3729  &0.0533                                \\
 0.050  & $     ^20^+$  &-14.4486  &0.1504 &                           & -14.3713  &0.0731                                \\
 0.100  & $     ^20^+$  &-14.4711  &0.2017 &-14.49603\footnotemark[2]  & -14.3698  &0.1004                                \\
 0.200  & $     ^20^+$  &-14.4653  &0.2614 &-14.49334\footnotemark[2]  & -14.3185  &0.1147                                \\
 0.500  & $  ^2(-1)^+$  &-14.3641  &0.3284 &-14.39454\footnotemark[2]  & -14.1517  &0.1159                                \\
 1.000  & $  ^2(-1)^+$  &-14.0674  &0.3154 &-14.07640\footnotemark[2]  & -13.8560  &0.1040                                \\
 2.000  & $  ^2(-1)^+$  &-13.2827  &0.2826 &                            & -13.092  &0.0916                                \\
 5.000  & $  ^2(-1)^+$  &-10.2740  &0.2556 &                            & -10.097  &0.0787                                \\
10.000  & $  ^2(-1)^+$  &-4.0679   &0.2593 &                            & -3.8927   &0.0840                               \\
\end{tabular}
\caption{Total energies E$_{\mbox{\scriptsize Tot}}$,  one-particle
  ionization energies E$_{\mbox{\scriptsize Ion}}$, and
previously published data E$_{\mbox{\scriptsize Lit}}$ in a. u. for the states $\nu^10^+$ ($\nu=1,2,3$), as well as the symmetries T$_{\mbox{\scriptsize{}Sym}}$ of the states at threshold
 for different field strengths $\gamma$.}\label{tab:1-1+} \footnotetext[1]{Ref. \cite{Galvez:2002_1}.} \footnotetext[2]{Ref. \cite{Guan:2003_1}.}
 
\end{ruledtabular}
\end{table*}

Let us turn to the discussion of our results for the triplet symmetry shown in
Fig.~\ref{fig:2S+1(-1)+}. The singlet and triplet symmetry subspaces are closely related.
Table \ref{tab:3-1+} shows, that the one-particle ionization threshold is
represented by the same states of the Be$^+$ ion. But the reader should keep in
mind, that the ionization thresholds for both subspaces do not possess the same energy,
because of the different spin projections, chosen for the total
energies. It can be seen in Fig.~\ref{fig:2S+1(-1)+}
that the ionization energies of the triplet states are typically larger 
than those of the corresponding singlet states. However, they
follow the same overall behavior: In the low field regime the energies
increase, whereas they decrease in the high field regime showing in between
a local maximum.  Furthermore one observes, that the OPIEs of the
$1^1(-1)^+$ and $1^3(-1)^+$ and the states $2^1(-1)^+$ and $2^3(-1)^+$ approach each other 
for the highest field strengths investigated here.

\begin{table*}[ht]
\begin{ruledtabular}
\begin{tabular}{*{3}{c}*{8}{c}}
\multicolumn{2}{c}{\rule[-0.7em]{0mm}{2.0em}} &\multicolumn{3}{c}{\rule[-0.7em]{0mm}{2.0em}{}$1^3(-1)^+$}& \multicolumn{3}{c}{$2 ^3(-1)^+$}& \multicolumn{2}{c}{$3 ^3(-1)^+$}\\
\hline
\multicolumn{1}{c}{\rule[-0.7em]{0mm}{2.0em}$\gamma $}&\multicolumn{1}{c}{T$_{\mbox{\scriptsize{}Sym}}$}  & \multicolumn{1}{c}{E$_{\mbox{\scriptsize{}tot}}$} &\multicolumn{1}{c}{E$_{\mbox{\scriptsize{}Ion}}$}  & \multicolumn{1}{c}{E$_{\mbox{\scriptsize{}Lit}}$}   & \multicolumn{1}{c}{E$_{\mbox{\scriptsize{}tot}}$}&  \multicolumn{1}{c}{E$_{\mbox{\scriptsize{}Ion}}$} &\multicolumn{1}{c}{E$_{\mbox{\scriptsize{}Lit}}$}& \multicolumn{1}{c}{E$_{\mbox{\scriptsize{}tot}}$}&  \multicolumn{1}{c}{E$_{\mbox{\scriptsize{}Ion}}$}\\
\hline
 0.000  & $     ^20^+$  &-14.5598  &0.2351   &-14.56724\footnotemark[1] & -14.3965  &0.0718& -14.39896\footnotemark[1] & -14.3543  &0.0296  \\
 0.001  & $     ^20^+$  &-14.5612  &0.2361   &-14.56388\footnotemark[2] & -14.3980  &0.0728&                           & -14.3581  &0.0329  \\
 0.010  & $     ^20^+$  &-14.5744  &0.2448   &-14.57721\footnotemark[2] & -14.4103  &0.0807&                           & -14.3642  &0.0346  \\
 0.050  & $     ^20^+$  &-14.6142  &0.2660  &-14.58281\footnotemark[3] & -14.4329  &0.0847 &                           & -14.3859  &0.0377  \\
 0.100  & $     ^20^+$  &-14.6936  &0.3242   &-14.69575\footnotemark[2] & -14.4787  &0.1093&                           & -14.4186  &0.0493  \\
 0.200  & $     ^20^+$  &-14.7979  &0.3941   &-14.80065\footnotemark[2] & -14.5156  &0.1118&                           & -14.4631  &0.0593  \\
 0.500  & $  ^2(-1)^+$  &-15.0107  &0.4749   &-15.01300\footnotemark[2] & -14.6616  &0.1258&                           & -14.5881  &0.0523  \\
 1.000  & $  ^2(-1)^+$  &-15.1982  &0.4462   &-15.19348\footnotemark[2] & -14.8751  &0.1231&                           & -14.8037  &0.0517  \\
 2.000  & $  ^2(-1)^+$  &-15.3551  &0.3551   &-15.30815\footnotemark[3] & -15.1036  &0.1035&                           & -15.0438  &0.0437  \\
 5.000  & $  ^2(-1)^+$  &-15.3002  &0.2818   &-15.25183\footnotemark[3] & -15.1011  &0.0827&                           & -15.0410  &0.0226  \\
10.000  & $  ^2(-1)^+$  &-14.0792  &0.2705   &-14.03046\footnotemark[3] & -13.8945  &0.0858&                           & -13.8320  &0.0233  \\
\end{tabular}
\caption{Total energies E$_{\mbox{\scriptsize Tot}}$,  one-particle
  ionization energies E$_{\mbox{\scriptsize Ion}}$, and
previously published data E$_{\mbox{\scriptsize Lit}}$ in a.u. for the states $\nu^3(-1)^+$ ($\nu=1,2,3$), as well as the symmetry T$_{\mbox{\scriptsize{}Sym}}$ of the ionic threshold states
at different field strengths 
$\gamma$.} \label{tab:3-1+}
 \footnotetext[1]{Ref. \cite{Galvez:2002_1}.} \footnotetext[2]{Ref. \cite{Guan:2003_1}.}
  \footnotetext[3]{Ref. \cite{Ivanov:2001_1}.} 
\end{ruledtabular}
\end{table*}

Values for total energies for finite $\gamma$ are available in the literature exclusively for the
$1^3(-1)^+$ state but not for the excitations of this symmetry. We therefore
discuss here only the accuracy of the energies of this state. In the absence of
the field our relative accuracy compared to the accurate results of Ref.\cite{Galvez:2002_1}
is approximately $5 \times 10^{-4}$. In the presence of a weak field our total energies are
above those of Ref.\cite{Guan:2003_1} by a relative difference of $2 \times 10^{-4}$.
This difference decreases with increasing field strength and for $\gamma > 0.5$
our results are variationally lower compared to those of Ref.\cite{Guan:2003_1}.
For $\gamma \ge 2$ only Hartree-Fock results are available \cite{Ivanov:2001_1}
which are systematically higher in energy by approximately $4 \times 10^{-4}$ relative
deviation.

\subsection{The symmetry subspaces $^3(-1)^{-}$}
\label{sec:3-1-}

We present in this subsection our results for the triplet states
for magnetic quantum number $M=-1$ and negative $z$ parity. The
ground and one excited state have been investigated here.
Although table \ref{tab:3-1-} shows, that the ionization thresholds for this
symmetry are the same as for the corresponding subspace with positive
$z$ parity, the ionization energies show a completely different
behavior. For the $1^3(-1)^{-}$ state it increases monotonically with
increasing field strength for the complete regime considered here:
from $0.0562$~a.u. at $\gamma=0$ to $0.5215$~a.u. at $\gamma=10$.
Comparing our results for the total energies of this state with the
literature we arrive at the following conclusions. For $\gamma = 0$
we have a relative accuracy of $10^{-3}$ compared to the values presented
in Ref.\cite{Galvez:2002_1}. In the presence of the field for $\gamma \le 0.5$  our
variational total energies are larger than those of Ref.\cite{Guan:2003_1} by
a relative change of $5 \times 10^{-4}$. For $\gamma = 1$ our value is below
that of Ref.\cite{Guan:2003_1} and for $\gamma > 1$ no data are available in
the literature for comparison. 

According to our study (and definition of the ionization threshold)
the $2^3(-1)^{-}$ state starts to possess a significant binding energy of the outer electron
from $\gamma \approx 0.2$ on.
Its ionization energy increases monotonically but the rate of this increase
decreases for $\gamma > 0.5$ and seems to approach an asymptotic value at approximately
$0.12$ a.u. No data are available in the literature on this state in the presence of
the field. In the absence of the field the total energy given in ref.\cite{Galvez:2002_1}
amounts to -14.38462 a.u.

\begin{figure}[htbp]
  \includegraphics*[scale=0.30]{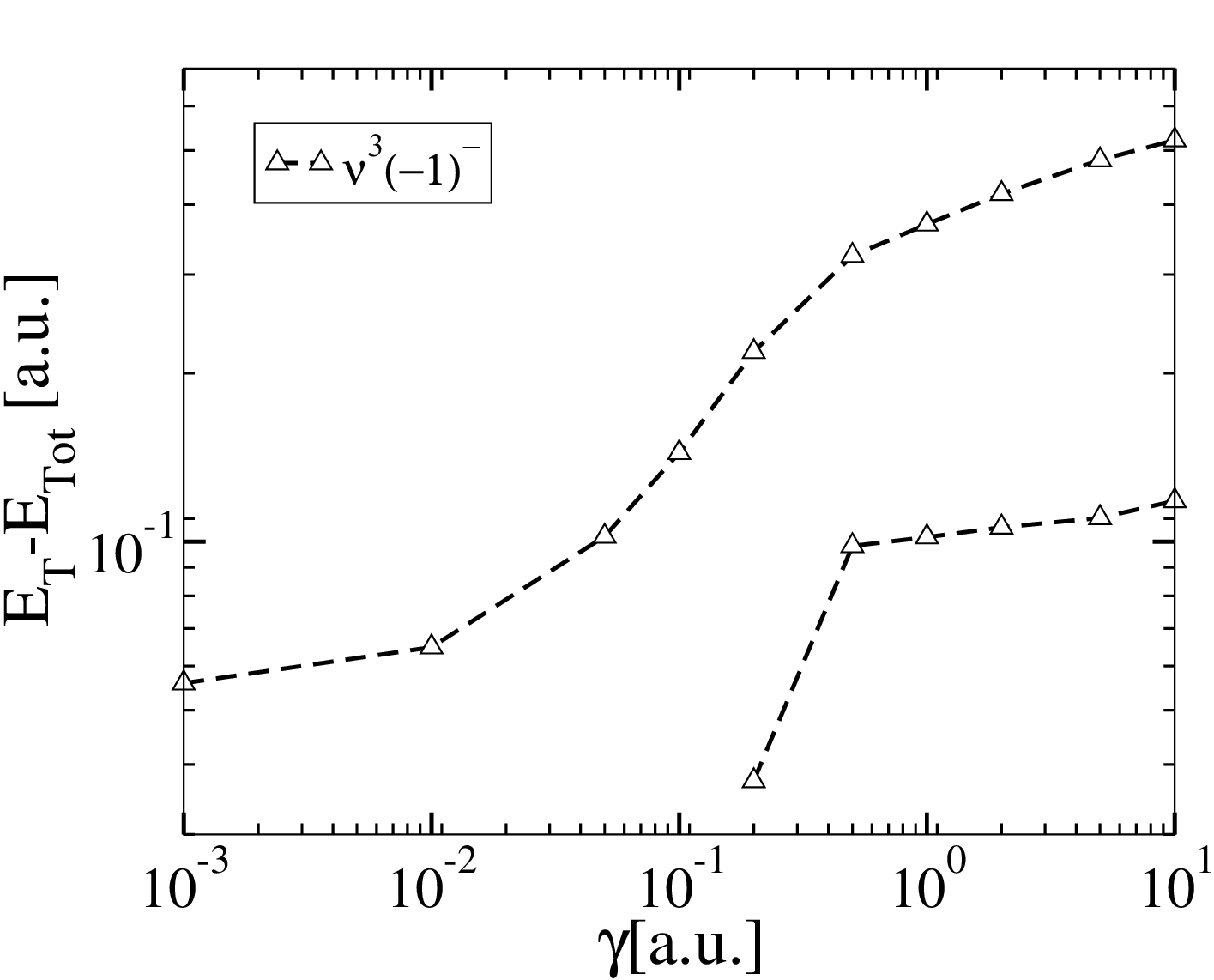}
  \caption{One-particle ionization energies for the states $\nu^3(-1)^-$ ($\nu=1,2$) in a.u. as a function of the magnetic field strength $\gamma$.}
  \label{fig:1(-1)-}
\end{figure}

\begin{table*}[ht]
\squeezetable
\begin{ruledtabular}
\begin{tabular}{*{3}{c}*{9}{c}}
\multicolumn{2}{c}{\rule[-0.7em]{0mm}{2.0em}} &\multicolumn{3}{c}{\rule[-0.7em]{0mm}{2.0em}{}$1^3(-1)^-$}& \multicolumn{2}{c}{$2 ^3(-1)^-$}& \multicolumn{3}{c}{$1 ^3(-3)^+$}& \multicolumn{2}{c}{$2 ^3(-3)^+$}\\
\hline\hline
\multicolumn{1}{c}{\rule[-0.7em]{0mm}{2.0em}$\gamma $}&\multicolumn{1}{c}{T$_{\mbox{\scriptsize{}Sym}}$}  & \multicolumn{1}{c}{E$_{\mbox{\scriptsize{}tot}}$} &\multicolumn{1}{c}{E$_{\mbox{\scriptsize{}Ion}}$}  & \multicolumn{1}{c}{E$_{\mbox{\scriptsize{}Lit}}$}   & \multicolumn{1}{c}{E$_{\mbox{\scriptsize{}tot}}$}&  \multicolumn{1}{c}{E$_{\mbox{\scriptsize{}Ion}}$}  & \multicolumn{1}{c}{E$_{\mbox{\scriptsize{}tot}}$} &\multicolumn{1}{c}{E$_{\mbox{\scriptsize{}Ion}}$}  & \multicolumn{1}{c}{E$_{\mbox{\scriptsize{}Lit}}$}   & \multicolumn{1}{c}{E$_{\mbox{\scriptsize{}tot}}$}&  \multicolumn{1}{c}{E$_{\mbox{\scriptsize{}Ion}}$}\\
\hline
 0.000  & $     ^20^+$  &-14.3809  &0.0562   &-14.39544\footnotemark[1]&             &          &            &            &                            &             &          \\
 0.001  & $     ^20^+$  &-14.3810  &0.0559   &-14.39066\footnotemark[2]&             &          &            &            &                            &             &          \\
 0.010  & $     ^20^+$  &-14.3944  &0.0648   &-14.40399\footnotemark[2]&             &          &            &            &                            &             &          \\
 0.050  & $     ^20^+$  &-14.4504  &0.1022   &                         &             &          &            &            &                            &             &          \\
 0.100  & $     ^20^+$  &-14.5136  &0.1442   &-14.52313\footnotemark[2]&             &          &-14.4082  &0.0388   &                            &             &          \\
 0.200  & $     ^20^+$  &-14.6223  &0.2184   &-14.63122\footnotemark[2]& -14.4412  &0.0374 &-14.5492  &0.1453   &                            &             &          \\
 0.500  & $  ^2(-1)^+$  &-14.8600  &0.3243   &-14.86797\footnotemark[2]& -14.6340  &0.0983 &-14.8530 &0.3172   &-14.82273\footnotemark[3]   & -14.6229  &0.0872 \\
 1.000  & $  ^2(-1)^+$  &-15.1210  &0.3690   &-15.11903\footnotemark[2]& -14.8539  &0.1019 &-15.1975  &0.4455   &-15.16179\footnotemark[3]   & -14.8605  &0.1085 \\
 2.000  & $  ^2(-1)^+$  &-15.4183  &0.4183   &                         & -15.1061  &0.1061 &-15.6217  &0.6216   &-15.57496\footnotemark[3]   & -15.1350  &0.1349 \\
 5.000  & $  ^2(-1)^+$  &-15.4990  &0.4805   &                         & -15.1287  &0.1103 &-15.9493  &0.9308   &-15.91027\footnotemark[3]   & -15.1674  &0.1490 \\
10.000  & $  ^2(-1)^+$  &-14.3302  &0.5215   &                         & -13.9269  &0.1183 &-15.0875  &1.2789   &-15.04644\footnotemark[3]   & -13.9344  &0.1257 \\
\end{tabular}
\caption{Total energies E$_{\mbox{\scriptsize Tot}}$,  one-particle
  ionization energies E$_{\mbox{\scriptsize Ion}}$, and
previously published data E$_{\mbox{\scriptsize Lit}}$ in a.u. for the states $\nu^3(-1)^-$ ($\nu=1,2$) and  $\nu^3(-3)^+$ ($\nu=1,2$) as well as the threshold symmetry T$_{\mbox{\scriptsize{}Sym}}$ at different field strengths 
$\gamma$.}\label{tab:3-1-} \footnotetext[1]{Ref. \cite{Galvez:2002_1}.} \footnotetext[2]{Ref. \cite{Guan:2003_1}.}\footnotetext[3]{Ref. \cite{Ivanov:2000_1}.}
 
\end{ruledtabular}
\end{table*}

\subsection{The $^{3/5}(-3)^+$ symmetry subspaces}
\label{sec:2S-3+}

Here we discuss the ground and a single excited state of spin triplet $^{3}(-3)^+$ symmetry
and the ground as well as two excited states of spin quintet $^{5}(-3)^+$ symmetry.

\begin{figure}[htbp]
  \includegraphics*[scale=0.30]{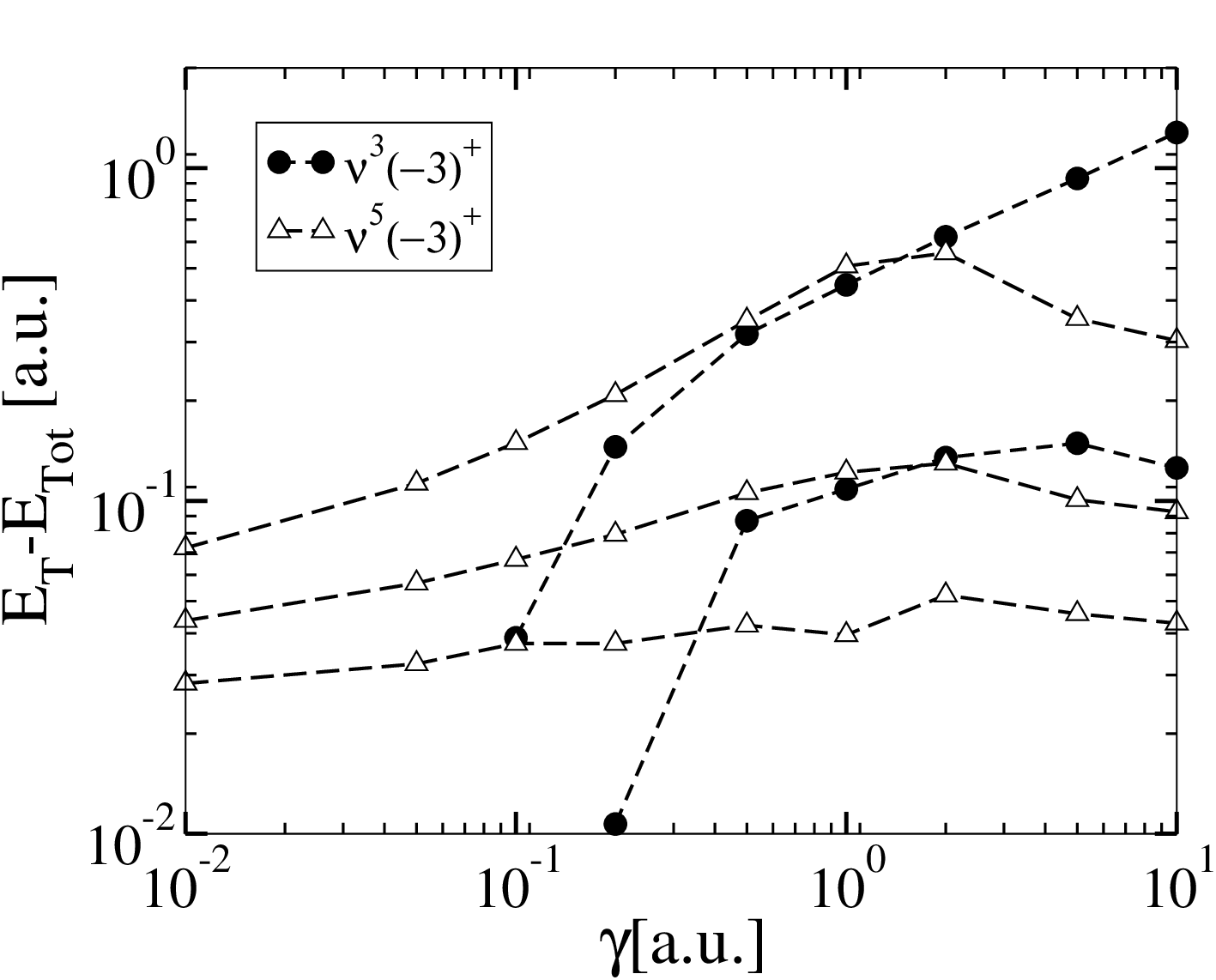}
  \caption{One-particle ionization energies for the states
    $\nu^{3}(-3)^+$ ($\nu=1,2$) and $\nu^{5}(-3)^+$ ($\nu=1,2,3$)  in
    a.u. as a function of the magnetic field strength $\gamma$.} 
  \label{fig:2S+1(-3)+}
\end{figure}

Let us begin our analysis with the spin triplet states.
Results for this symmetry subspace are presented in Fig.~\ref{fig:2S+1(-3)+}
and table \ref{tab:3-1-}. The ground state of the spin triplet symmetry with
magnetic quantum number $-3$ and positive $z$ parity is a quadruply tightly
bound state, being predominantly described by the orbitals $1s^22p_{-1}3d_{-2}$. 
These orbitals are the energetically lowest tightly bound orbitals.
As a consequence the $1^3(-3)^+$ state becomes for sufficiently high field strengths the state
with the largest OPIE among the states considered here (the most tightly
bound state of the beryllium atom in the high field limit is the $1s^2 2p_{-1}^2$-state which 
however is a spin singlet state and is not investigated here).
This holds particularly in the high
field limit, although in this field regime the total energy of the
quintet states is much lower. For $\gamma=0$ the ionization energy amounts to
$0.0388$~a.u. and at $\gamma=10$ to $1.2789$~a.u. The corresponding monotonous 
increase of the OPIE can clearly be seen in Fig.~\ref{fig:2S+1(-3)+}.
For field strengths $\gamma < 0.2$ we could not detect the existence of this
bound state in the presence of the magnetic field. For $\gamma \ge 0.5$ the 
only available values for the total energies in the literature
are due to the Hartree-Fock approach employed in Ref.\cite{Ivanov:2000_1}.
Our values are systematically lower than the Hartree-Fock values showing
a typical relative deviation of $3 \times 10^{-3}$.

The ionization energy of the first excited state $2^3(-3)^+$ increases
monotonically for $\gamma < 5$ but shows a decrease when doubling the
field strength to $\gamma = 10$. We obtain an ionization energy 
$0.0872$~a.u. at $\gamma=0.5$ and $0.1490$~a.u. for $\gamma=5$.
This state has up to date not been investigated in the literature.

Let us now discuss the states of the quintet spin symmetry.
The ground state $1^5(-3)^+$ of this symmetry is a triply tightly bound 
state, which is predominantly described by the configuration
$1s2s2p_{-1}3d_{-2}$. A monotonically increasing
OPIE for $\gamma<2$ can be observed for this state
(see Fig.~\ref{fig:2S+1(-3)+}). For $\gamma \geq 2$ the ionization energy decreases,
which is a consequence of the change with respect to the ionization threshold:
For lower fields the ionization threshold involves the doubly tightly bound
$1^4(-1)^+$ state of Be$^+$ and for $\gamma \geq 2$ it involves the triply tightly bound state
$1^4(-3)^+$. Similar statements hold for the first and second
excited states of this symmetry. Table \ref{tab:5-3+} shows, that the increase of
the OPIE for the ground state $1^5(-3)^+$
as well as the first and second excited state $2^5(-3)^+$ and $3^5(-3)^+$
are very strong for $\gamma<2$. For the ground state the 
ionization energy increases from $0.0547$~a.u. at $\gamma=0$ to
$0.5558$~a.u. at $\gamma=2$, i.e. by approximately one order of magnitude.
For the first and second excited state we encounter
$0.0293$~a.u. and $0.0172$~a.u. at $\gamma=0$ and
$0.1297$~a.u. and $0.0521$~a.u., respectively, at $\gamma=2$, i.e. an increase by approximately
a factor of 4 respectively 3. In the regime $\gamma \geq 2$ the relative decrease of
the ionization energy of the ground state is stronger than those for the first 
and the second excited states. Data from the literature to compare with
are only available for the ground state $1^5(-3)^+$ of this symmetry and only
Hartree-Fock results in the strong field regime.
Comparing with the latter data \cite{Ivanov:2000_1}
our total energies are systematically lower by a relative change of $5 \times 10^{-4}$.

\begin{table*}[ht]\squeezetable
\begin{ruledtabular}
\begin{tabular}{*{3}{c}*{6}{c}}
\multicolumn{2}{c}{\rule[-0.7em]{0mm}{2.0em}} &\multicolumn{3}{c}{\rule[-0.7em]{0mm}{2.0em}{}$1^5(-3)^+$}& \multicolumn{2}{c}{$2 ^5(-3)^+$}& \multicolumn{2}{c}{$3 ^5(-3)^+$}\\
\hline\hline
\multicolumn{1}{c}{\rule[-0.7em]{0mm}{2.0em}$\gamma $}&\multicolumn{1}{c}{T$_{\mbox{\scriptsize{}Sym}}$}  & \multicolumn{1}{c}{E$_{\mbox{\scriptsize{}tot}}$} &\multicolumn{1}{c}{E$_{\mbox{\scriptsize{}Ion}}$}  & \multicolumn{1}{c}{E$_{\mbox{\scriptsize{}Lit}}$}   & \multicolumn{1}{c}{E$_{\mbox{\scriptsize{}tot}}$}&  \multicolumn{1}{c}{E$_{\mbox{\scriptsize{}Ion}}$}  & \multicolumn{1}{c}{E$_{\mbox{\scriptsize{}tot}}$} &\multicolumn{1}{c}{E$_{\mbox{\scriptsize{}Ion}}$}     \\
\hline
 0.000  & $  ^4(-1)^+$  & -10.1197 & 0.0547   &                          & -10.0943 & 0.0293   & -10.0822   &0.0172 \\
 0.010  & $  ^4(-1)^+$  & -10.1245 & 0.0590   &                          & -10.0991 & 0.0336   & -10.0870   &0.0215 \\
 0.010  & $  ^4(-1)^+$  & -10.1549 & 0.0722   &                          & -10.1265 & 0.0437  &  -10.1111  & 0.0283\\
 0.050  & $  ^4(-1)^+$  & -10.2736 & 0.1129   &                          & -10.2173 &0.0566  &  -10.1930  & 0.0324 \\
 0.100  & $  ^4(-1)^+$  &-10.4070  &0.1495   &                           & -10.3243 &0.0667  &  -10.2948  & 0.0373 \\
 0.200  & $  ^4(-1)^+$  & -10.6438 & 0.2085   &                          & -10.5146 &0.0793  &  -10.4726  & 0.0373 \\
 0.500  & $  ^4(-1)^+$  & -11.2406 &0.3487    &-11.23262\footnotemark[1] & -10.9975 &0.1057  &  -10.9342  & 0.0521  \\
 1.000  & $  ^4(-1)^+$  & -12.0042 & 0.5076   &-11.99646\footnotemark[1] & -11.6186 &0.1219  &  -11.5363  & 0.0396  \\
 2.000  & $  ^4(-3)^+$  &-13.1560  &0.5558   &-13.14233\footnotemark[1]  & -12.7300 & 0.1297 &  -12.6523  & 0.0521\\
 5.000  & $  ^4(-3)^+$  & -15.7897 &  0.3529  &-15.78294\footnotemark[1] & -15.5375 & 0.1008 &  -15.4826  & 0.0458\\
10.000  &  $  ^4(-3)^+$ & -19.1314 & 0.3030   &-19.12479\footnotemark[1] & -18.9211 & 0.0927 &  -18.8712  & 0.0429\\
\end{tabular}
\caption{Total energies E$_{\mbox{\scriptsize Tot}}$,  one-particle
  ionization energies E$_{\mbox{\scriptsize Ion}}$, and
previously published data E$_{\mbox{\scriptsize Lit}}$ in a. u. for the states $\nu^5(-3)^+$ ($\nu=1,2,3$)  as well as the threshold symmetry T$_{\mbox{\scriptsize{}Sym}}$ at different field strengths 
$\gamma$.}\label{tab:5-3+} \footnotetext[1]{Ref. \cite{Ivanov:2000_1}.}
 
\end{ruledtabular}
\end{table*}

\subsection{The $^{5}(-3)^-$ and $^5(-6)^+$ symmetry subspaces}
\label{sec:5-3-}

\begin{figure}[htbp]
  \includegraphics*[scale=0.30]{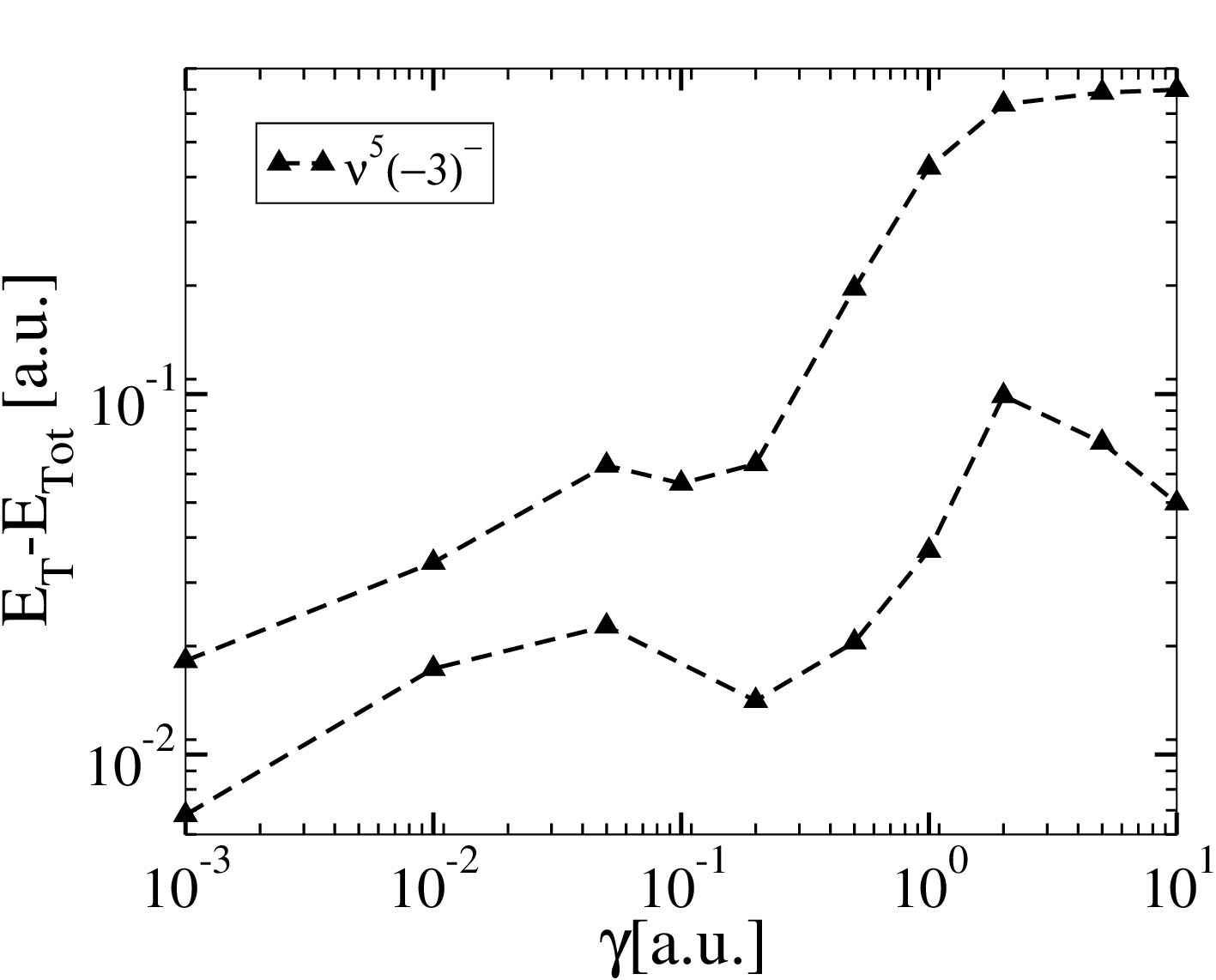}
  \caption{One-particle ionization energies for the states $\nu^5(-3)^-$ ($\nu=1,2$) in a.u. as a function of the magnetic field strength $\gamma$.}
  \label{fig:5(-3)-}
\end{figure}

In this subsection we discuss the spin quintet states with magnetic quantum numbers
$-3$ and $-6$ possessing negative and positive $z-$parity, respectively.
For the states of $^{5}(-3)^-$ symmetry a particular behavior of their ionization
energies  can be observed in Fig.~\ref{fig:5(-3)-}. The corresponding numerical results
are given in table \ref{tab:5-3-}. For the ground state the
OPIE increases in the field regime
$0<\gamma<0.05$: we encounter $0.0172$~a.u. at $\gamma=0.001$ and
$0.0572$~a.u. at $\gamma=0.05$. In the regime $0.05<\gamma<0.2$ it passes through
a local minimum and for $0.2<\gamma<2$ a rapid increase of the ionization
energy with increasing field strength takes place. At $\gamma=0.2$ it amounts to $0.0578$~a.u.
whereas at $\gamma=2$ it is one order of magnitude larger and amounts to $0.6364$~a.u. 
Thereafter for even larger field strengths the increase is slowing down
significantly. The reason for the latter is a change with respect to the
ionization threshold. For $\gamma \le 1$ the ionization threshold involves
the Be$^+$ state $^4(-1)^+$ where as for $\gamma \gtrsim 1$ 
the triply tightly bound Be$^+$ state $1^4(-3)^+$ defines the threshold. It represents the high
field global ground state of the beryllium positive ion.

The OPIE of the first excited state $2^5(-3)^-$ shows a similar behavior as
that of the ground state. However, in contrast to the 
ground state its ionization energy decreases for $\gamma \geq 2$ significantly. 
Again only for the ground state $1^5(-3)^-$ and in the strong field regime there
are data of Hartree-Fock calculations \cite{Ivanov:2000_1} available to compare with. Equally 
our total energies are variationally lower than those of the Hartree-Fock calculations
by a relative deviation of $10^{-3}$. 

\begin{table*}[ht]
\begin{ruledtabular}
\begin{tabular}{*{3}{c}*{9}{c}}
\multicolumn{2}{c}{\rule[-0.7em]{0mm}{2.0em}} &\multicolumn{3}{c}{\rule[-0.7em]{0mm}{2.0em}{}$1^5(-3)^-$}& \multicolumn{2}{c}{$2 ^5(-3)^-$}& \multicolumn{3}{c}{$1 ^5(-6)^+$}& \multicolumn{2}{c}{$2 ^5(-6)^+$}\\
\hline
\multicolumn{1}{c}{\rule[-0.7em]{0mm}{2.0em}$\gamma $}&\multicolumn{1}{c}{T$_{\mbox{\scriptsize{}Sym}}$}  & \multicolumn{1}{c}{E$_{\mbox{\scriptsize{}tot}}$} &\multicolumn{1}{c}{E$_{\mbox{\scriptsize{}Ion}}$}  & \multicolumn{1}{c}{E$_{\mbox{\scriptsize{}Lit}}$}   & \multicolumn{1}{c}{E$_{\mbox{\scriptsize{}tot}}$}&  \multicolumn{1}{c}{E$_{\mbox{\scriptsize{}Ion}}$}  & \multicolumn{1}{c}{E$_{\mbox{\scriptsize{}tot}}$} &\multicolumn{1}{c}{E$_{\mbox{\scriptsize{}Ion}}$}  & \multicolumn{1}{c}{E$_{\mbox{\scriptsize{}Lit}}$}   & \multicolumn{1}{c}{E$_{\mbox{\scriptsize{}tot}}$}&  \multicolumn{1}{c}{E$_{\mbox{\scriptsize{}Ion}}$}\\
\hline
 0.001  & $  ^4(-1)^+$  &-10.0826  &0.0172   &                & -10.0712 &0.0057 \\
 0.010  & $  ^4(-1)^+$  &-10.1107  &0.0279   &                & -10.0939 &0.0112 \\
 0.050  & $  ^4(-1)^+$  &-10.2179  &0.0572   &                & -10.1773 &0.0166 \\
 0.100  & $  ^4(-1)^+$  &-10.3079  &0.0503   &                                        \\
 0.200  & $  ^4(-1)^+$  &-10.4931  &0.0578   &                & -10.4431 &0.0078 \\
 0.500  & $  ^4(-1)^+$  &-11.0818  &0.1899   &-11.06254\footnotemark[1] & -10.9058  &0.0139 \\
 1.000  & $  ^4(-1)^+$  &-11.9151  &0.4184   &-11.89891\footnotemark[1] & -11.5260  &0.0294 &-11.7358  &0.2389      &-11.72880\footnotemark[1]   &          &                           \\       
 2.000  & $  ^4(-3)^+$  &-13.2366  &0.6364   &-13.22133\footnotemark[1] & -12.6992  &0.0989 &-13.1762  &0.5759      &-13.16961\footnotemark[1]   & -12.7321  &0.1319     \\
 5.000  & $  ^4(-3)^+$  &-16.1233  &0.6866   &-16.10812\footnotemark[1] & -15.5101  &0.0734 &-16.3139  &0.8772      &-16.30690\footnotemark[1]   & -15.5887  &0.1520     \\
10.000  & $  ^4(-3)^+$  &-19.5270  &0.6987   &-19.51207\footnotemark[1] & -18.8781  &0.0498 &-20.0242  &1.1959      &-20.01753\footnotemark[1]   & -18.9887  &0.1604     \\
\end{tabular}
\caption{Total energies E$_{\mbox{\scriptsize Tot}}$,  one-particle
  ionization energies E$_{\mbox{\scriptsize Ion}}$, and
previously published data E$_{\mbox{\scriptsize Lit}}$ in a.u. for the states $\nu^5(-3)^-$ ($\nu=1,2$) and  $\nu^5(-6)^+$ ($\nu=1,2$) as well as the threshold symmetry T$_{\mbox{\scriptsize{}Sym}}$ at different field strengths 
$\gamma$.}\label{tab:5-3-}\footnotetext[1]{Ref. \cite{Ivanov:2000_1}.}
 
\end{ruledtabular}
\end{table*}

Let us now turn to the states of $^{5}(-6)^+$ symmetry.
The ground state $1^{5}(-6)^+$ represents the global
ground state of the beryllium atom in the high field limit. It is
the energetically lowest (we are refering, of course, to the total energy)
quadruply tightly bound quintet state and is predominately described by the configuration
$1s2p_{-1}3d_{-2}4f_{-3}$. However, this configuration does not represent the most tightly
bound state with four occupied tightly bound orbitals for any field strength.
The OPIEs of the ground and the first excited state increase monotonically in the strong field
regime $\gamma\geq1$. In the low field regime these states represent highly excited states
in the continuum which are not bound. Similar to the previously discussed cases also
for this symmetry there are only Hartree-Fock data \cite{Ivanov:2000_1} available
for the ground state. Comparing with the latter our total energies are variationally lower
by typically $4 \times 10^{-4}$.

\begin{figure}[htbp]
  \includegraphics*[scale=0.30]{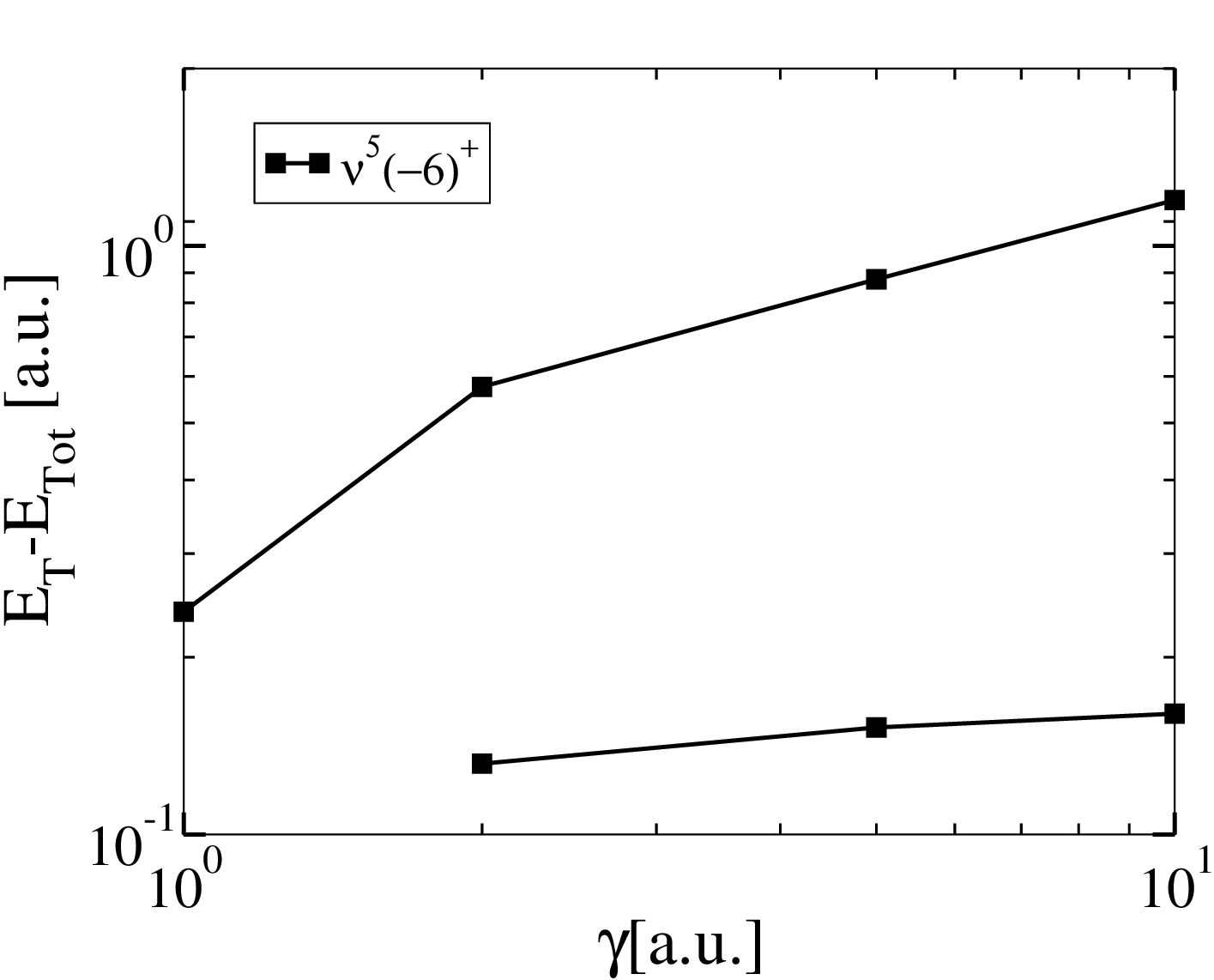}
  \caption{One-particle ionization energies for the states $\nu^5(-6)^-$ ($\nu=1,2$)
  in a.u. as a function of the magnetic field strength $\gamma$.}
  \label{fig:5(-6)+}
\end{figure}

\section{Electromagnetic transitions}
\label{sec:wave}

\begin{figure}[htbp]
  \centering
  \includegraphics*[scale=0.3]{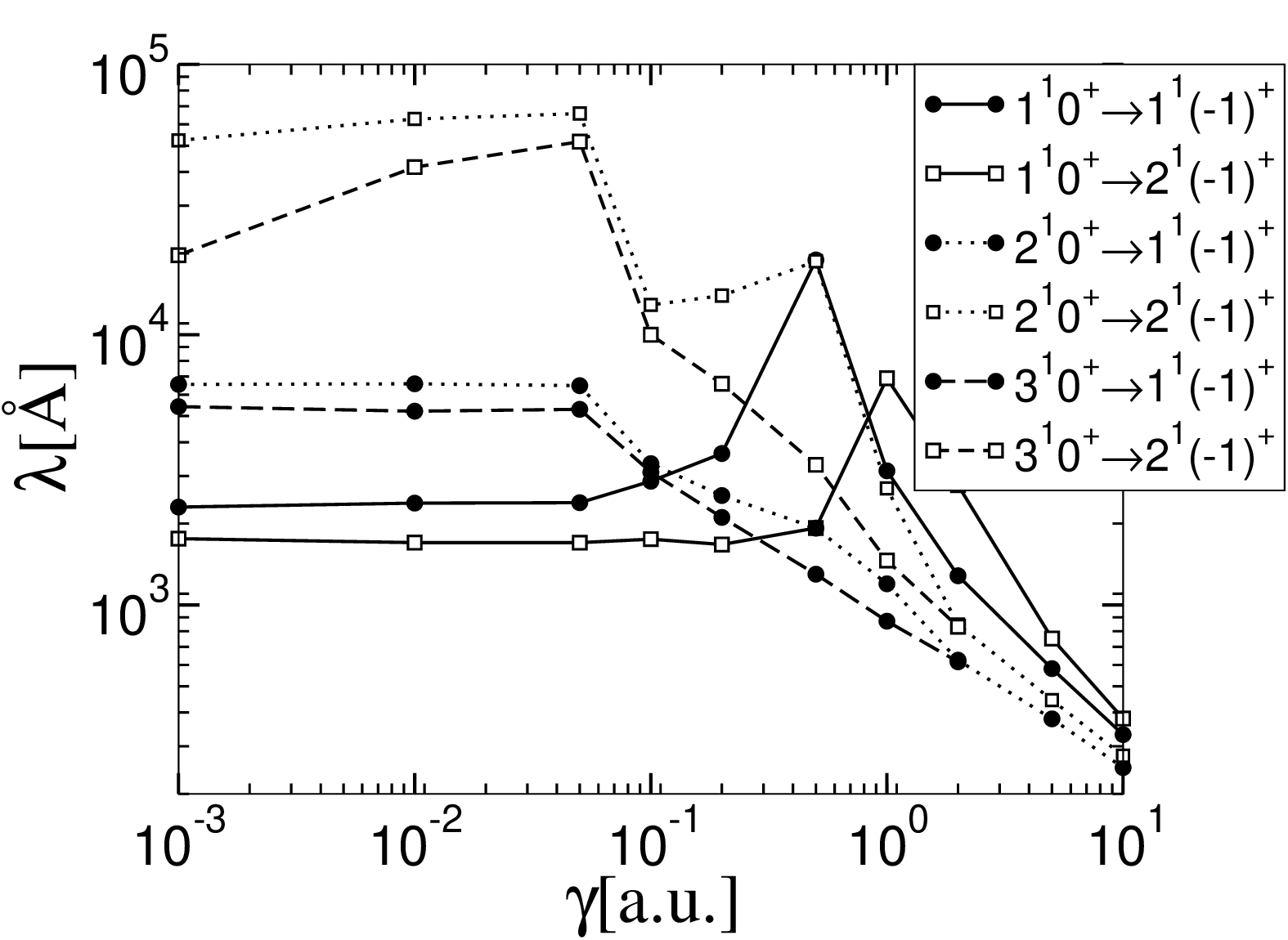}
  \caption{Transition wavelengths $\lambda$ for the circular polarized
  transitions $\nu^10^+\longrightarrow\mu^1(-1)^+$ ($\nu=1,2,3, \mu=1,2$)
  in \AA{} as a function of the magnetic field strength $\gamma$.}
  \label{fig:be_trans_0_1_-1_1}
\end{figure}

\begin{figure}[htbp]
  \centering
  \includegraphics*[scale=0.3]{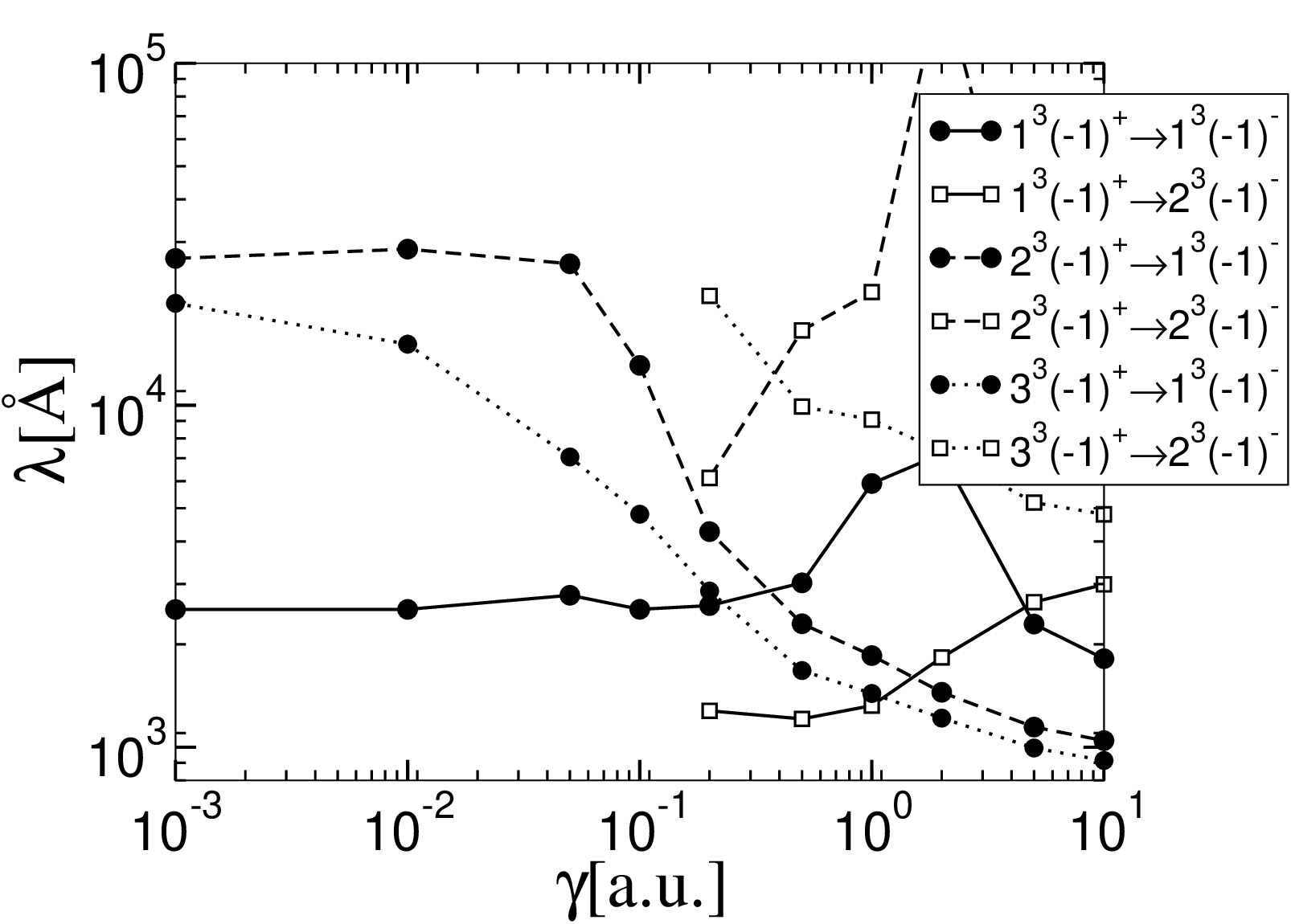}
  \caption{Transition wavelengths $\lambda$ for the linear polarized
  transitions $\nu^3(-1)^+\longrightarrow\mu^3(-1)^-$ ($\nu=1,2,3, \mu=1,2$)
  in \AA{} as a function of the magnetic field strength $\gamma$.
  For $\gamma > 0.05$ the reader observes that the wavelengths corresponding
  to the transitions with $(\nu,\mu)=(2,1),(3,1)$ and $(3,2)$ decrease monotonically
  whereas those for the transitions $(1,1),(1,2)$ and $(2,2)$ exhibit distinct
  maxima.}
  \label{fig:be_trans_-1_3+_-1_3-}
\end{figure}

In the present section we discuss the behavior of the wavelengths
$\lambda$ of the allowed electric dipole transitions as a function
of the magnetic field strength. The regime of wavelengths to be
considered is $\lambda<10^5$~\AA{}. We will focus on the
circular polarized transitions $\nu^10^+\longrightarrow\mu^1(-1)^+$
($\nu=1,2,3, \mu=1,2$) shown in Fig.~\ref{fig:be_trans_0_1_-1_1} and
the linear polarized transitions 
$\nu^3(-1)^+\longrightarrow\mu^3(-1)^-$ ($\nu=1,2,3, \mu=1,2$) presented
in Fig. \ref{fig:be_trans_-1_3+_-1_3-}. 

For the circular polarized transitions $\nu^10^+\longrightarrow\mu^1(-1)^+$
($\nu=1,2,3, \mu=1,2$) presented in Fig.~\ref{fig:be_trans_0_1_-1_1}
one observes, that in the field range $0<\gamma<0.05$ the
transition wavelengths are approximately constant. In this field regime the
influence of the magnetic field is weak. However, in the intermediate and high
field regime the transition wavelengths show a pronounced
field-dependence. For $\gamma>1$ an overall
decrease of the transition wavelengths is observable. In the regime
$0.05<\gamma<1$ some lines increase, others do not.
To be more specific: The transitions $(\nu,\mu)=(1,1),(1,2),$ and 
$(2,2)$ show an increase of the wavelengths, whereas the transitions $(\nu,\mu)=(2,1),
(3,1),$ and $(3,2)$ show a monotonous decrease. The origin of this behavior is the
less rapid increase of the total energies of the $\mu^1(-1)^+$ states
compared to the $\nu^10^+$ states. Therefore crossovers of the corresponding energy
levels occur, which lead to divergent transition wavelengths (these are, due to
the crude grid of field strengths employed here, visible as peaks in the 
figure ~\ref{fig:be_trans_0_1_-1_1}).

For the linear polarized transitions
$\nu^3(-1)^+\longrightarrow\mu^3(-1)^-$ ($\nu=1,2,3, \mu=1,2$) depicted 
in Fig.~\ref{fig:be_trans_-1_3+_-1_3-}, the behavior in the low field
regime $\gamma<0.05$ is very similar to the one of the circular polarized transitions discussed
above, i.e. two of the three wavelengths in this regime are almost independent of the
field strength. This does not hold for the transition
$3^3(-1)^+\longrightarrow1^3(-1)^-$ (shown with dotted line and black
circles). The latter is due to the fact, that the OPIE 
of the state $3^3(-1)^+$ is nearly field independent and is at
$\gamma=0$ only a factor of two smaller than the OPIE
of the state $1^3(-1)^-$. The ionization energy of
the latter state increases for weak fields ($\gamma<0.05$) about a factor of
two and as a result the corresponding transition wavelengths decrease.

\section{Conclusions}
\label{sec:summary}

We have investigated the electronic structure of the beryllium atom exposed to a strong 
magnetic field in the regime $0 \le \gamma \le 10$ a.u. Our approach is based on a recently
developed full configuration interaction method for multi-electron systems that employs
anisotropic Gaussian orbitals which are optimized for each field strength separately.
Our results comprise the global ground state and its crossovers as a function of the
field strengths. Earlier results on the Hartree-Fock level have been confirmed and refined.
Beyond this we have studied many excited states of various symmetries thereby covering
spin singlet, triplet and quintet states with magnetic quantum numbers $|M|=0,1,3,6$
and for both positive and negative $z-$parity. A total of 18 excited states (not counting
the spin multiplicity) have been studied
and analyzed i.e. their total and one-particle ionization energies as well as the 
corresponding transition wavelengths for allowed dipole transitions have been discussed.
The one-particle ionization energies show a rich variability as a function of the field
strength depending in particular on the number and type of tightly bound orbitals that
are contained in the configurations dominating the individual electronic states.
Many excited states have been investigated for the first time for finite field strengths.
For the states that have already been studied in the literature a comparison of our data
shows that our total energies are variationally higher by a relative change
of typically $10^{-3}$ or less for the regime of field strengths $0\le \gamma \le 0.5$.
This discrepancy is however not due to limitations of our basis set,
which possesses a very high flexibility,
but more an effect of the large number of configurations emerging for a four-electron system
within a full configuration interaction approach for a given (moderately sized) basis set.
Going to larger atomic systems it is 
therefore desirable to choose another configuration interaction approach instead of the
full one which allows to treat larger basis sets. For $\gamma \ge 0.5$ our results are almost
exclusively variationally lower than the existing ones in the literature by typical relative
deviations of several times $10^{-3}$. This confirms the already known feature
that our approach performs best in the intermediate regime of field strengths.

\section{Acknowledgments}
P.S. acknowledges fruitful discussions with J. Liebert.

\end{document}